\documentclass[%
 reprint,
nofootinbib,
 amsmath,amssymb,
 aps,
]{revtex4-2}

\usepackage{graphicx}
\usepackage{dcolumn}
\usepackage{bm}
\newcommand{\vn}[1]{\smash{#1}\vphantom{X^i_j}}

\begin{document}

\preprint{APS/123-QED}

\title{Viability Space Decomposition: \\ A geometric partition of survival outcomes in single- and multi-agent systems}
\thanks{By ``agents" we are broadly referring to organisms. This phrasing is typical within the history of viability theory.}%

\author{Connor McShaffrey}
 \email{cmcshaff@iu.edu}
\author{Randall D. Beer}%
 \altaffiliation[Also in the ]{Program in Neuroscience and Informatics Department, Indiana University Bloomington.}
\affiliation{%
Cognitive Science Program, Indiana University Bloomington\\
}%

\date{\today}

\begin{abstract}
What determines whether an organism or collective will survive under particular conditions? This question is asked across the life sciences when determining adaptive fit, developing efficacious treatments for diseases, and assessing the risks posed by ecological shifts. To aid their investigations, researchers employ models of agents which must respect particular constraints to remain alive. By constraining the dynamics of these agents to bounded viability regions, these models form a class of extended dynamical systems where transient dynamics can lead to death, making traditional attractors and separatrices insufficient for characterizing the global space of possible behaviors. To remedy this, we develop \textit{viability space decomposition}, an analysis framework for ordinary differential equation models of agents with viability constraints. We first introduce the general theory, revealing how several new classes of manifolds (mortality, ordering, and collapse) permit a complete decomposition of state space into regions of qualitatively similar survival outcomes: a \textit{viability portrait}. We then demonstrate the method by completely analyzing the global behavior of three models: a subcellular network, a behaving cell with the same physiology, and two coupled cell networks. Finally, we finish by discussing how the framework scales and future directions for its development and application.  
\end{abstract}

\maketitle


\section{Introduction}

One of life's defining features is its ability to defy --- to maintain its precarious existence in the face of forces that would otherwise erase it \cite{schrodingerWhatLifePhysical1992,phillipsSchrodingersWhatLife2021,beerTheoreticalFoundationsEnaction2023}. Whether it be designing safe and efficacious treatments for diseases, determining which conditions lead a cell to properly commit to death, or exploring whether a species will be able to adapt to an ecological shift, throughout the life sciences we are forced to grapple with how organisms' dynamics unfold relative to their constraints on existence. Yet, despite this ubiquity, we currently lack the theoretical machinery to have a principled understanding of this process. 

Currently, some of the best tools we have for constructing rigorous theories in the behavioral and life sciences are our mathematical and computational models. Agent-focused models are particularly powerful as they allow us to explore phenomena at the resolution of the participating individual(s). This level of detail allows us to explore how the fates and behaviors of individuals shape collective phenomena, and the impact this broader context recursively has on the individual, but it also makes these systems some of the most challenging to analyze.  Unlike population-level models where death is summarized by a simple rate \cite{loreauPopulationsEcosystemsTheoretical2010}, models with an agential perspective need to specify detailed conditions for individuals' deaths in order to make predictions about the connection between states and eventual fates \cite{mcshaffreyMattersLifeDeath2026}. This is as true for modeling physiological subsystems as it is for agent-based models of embodied and interacting individuals \cite{johLyseNotLyse2011,ghaffarizadehPhysiCellOpenSource2018}. While imposing such constraints on our models is not very mathematically demanding, and can even take the form of a simple threshold, understanding how the dynamics interact with these constraints is more elusive. As a result, the majority of model analyses rely on running a vast number of simulations and tracking their resulting outcomes without strong guiding principles for how the global space of possible solutions might be structured \cite{mcshaffreyMaintainingViabilityMultiple2022}. 

Relative to the ubiquity of this issue of how agents' dynamics unfold relative to their limits, a comparatively smaller body of work has focused on the theoretical question of how we can effectively analyze such systems. This endeavor, broadly encapsulated under the name \textit{viability theory}, can be traced back to the work of cybernetician Ross Ashby, who explicitly described physiological limits as forming a boundary around a compact subset of the system's state space where the organism is considered alive \cite{ashbyPhysicalOriginAdaptation1945,ashbyDesignBrain1960}. Since then, similar geometric perspectives on the life-death boundary have been applied in the study of adaptive behavior \cite{beerDynamicalSystemsPerspective1995}, cognitive science \cite{virgoGoodRegulatorTheorem2025}, personalized medicine \cite{voitSystemstheoreticalFrameworkHealth2009,davisDynamicalSystemsApproaches2019}, and artificial life \cite{barandiaranNormEstablishingNormFollowingAutonomous2014}, but very few of these works developed the theoretical tools needed for analysis. Some notable exceptions include frameworks which developed methods for control theoretic systems \cite{aubinViabilityTheoryNew2011,himeokaTheoreticalBasisCell2024}, assessing the ``normative" behavior of an agent \cite{barandiaranNormEstablishingNormFollowingAutonomous2014}, and information theoretic approaches \cite{egbertMethodsMeasuringViability2018,kolchinskySemanticInformationAutonomous2018,bartlettPhysicsLifeExploring2025}. Notably, though, all of these current approaches have one or more of the following limitations: First, not all of them seek to describe global organizing structure for viability, and instead still rely on having to sample many points throughout state space. Second, for the ones that do compute such global structures, it is not always clear that the principles can be rigorously scaled to arbitrary dimensions. Third, to the best of our knowledge, none of these frameworks are currently specified to deal with the integrity of more than one agent or system, preventing their application to most natural phenomena. 

In response, our goal in this paper is to present a theoretical approach for viability for one class of models, ordinary differential equations with physiological constraints, which overcomes all three of the aforementioned limitations. More specifically, we present \textit{viability space decomposition}, a framework which builds upon the geometric perspective by identifying novel global manifolds which separate regions of state space based on the existential outcomes they lead to, culminating in a \textit{viability portrait}. We start by presenting the general framework for single- and multi-agent systems, building on the dynamical approach to agent-environment interaction. From there, we present three models and use the approach to completely analyze their possible survival outcomes as worked examples. To conclude, we discuss the potential applications of the framework, its limitations, and future research directions. 

\section{Dynamics of Agent-Environment Interaction}

Within the life and behavioral sciences, there has been a strong push to develop a formal mathematical framework that encompasses the behavior of agents \cite{baltieriMathematicalApproachesStudy2025}. In this paper, we will be building on one well-known proposal which seeks to understand the reciprocal relationship between an agent and its environment, including other agents, through the lens of dynamical systems theory \cite{beerDynamicalSystemsPerspective1995}. The general idea is that, if we look at agents and the environment individually, each of them can be seen as a nonautonomous dynamical system which takes time-varying inputs from the other model components. Taken together, the components form an autonomous dynamical system in which the whole is fully determined. Most commonly, this perspective has been associated with the study of how the joint dynamics of brain, body, and environment give rise to particular behaviors \cite{chielBrainHasBody1997}, but in principle any partition is possible, regardless of whether the agents have nervous systems. For the purposes of this paper, we will simply partition the system into individual agents and their shared environment.  

Starting with a set of agents \(\alpha=\{\alpha_1,\ldots,\alpha_N\}\) and an environment \(\mathcal{E}\), we will assume that each of these systems has a state space \(X\) equivalent to \(\mathbb{R}^n\). The overall state space of the model universe \(\mathcal{U}\) can then be taken to be the Cartesian product of these state spaces: 
\begin{equation}
\mathcal{U}= \left( \prod_{i=1}^{N} X_{\alpha_i} \right) \times X_\mathcal{E}
\end{equation}
The dynamics of these partitioned subsystems are then determined by a set of, usually nonlinear, differential equations, designated by \(\dot{\mathbf{x}}_{\alpha_i}\) and \(\dot{\mathbf{x}}_{\mathcal{E}}\). This set of differential equations can be composed into a vector-valued function \(\mathbf{F}(\mathbf{x})\) that establishes a vector field over \(\mathcal{U}\), defining the overall rate of change \(\dot{\mathbf{x}}\) at each point (note that if the environment is treated as constant, the \(\mathcal{E}\) variables drop out of this equation, and state space will only be comprised of variables belonging to agents):
\begin{equation}
\dot{\mathbf{x}} = \begin{bmatrix}
           \dot{\mathbf{x}}_{\alpha_1} \\
           \vdots \\
           \dot{\mathbf{x}}_{\alpha_N} \\
           \dot{\mathbf{x}}_{\mathcal{E}}
         \end{bmatrix} = \mathbf{F}(\mathbf{x})
\end{equation}
This vector field over \(\mathcal{U}\) implicitly defines a global flow \(\phi_t\), telling us how states will evolve over time, including from particular initial condition \(\phi_t(\mathbf{x}_0)\). By identifying the limit sets and corresponding invariant manifolds that organize the flow, dynamical systems theory is capable of summarizing global behavior in a  \textit{phase portrait}, which is typically a complete analysis. Nonetheless, this theoretical machinery alone is not sufficient for analyzing models where agents can perish. In the next section, we explore how the addition of viability constraints complicates this picture. 

\section{The Viability Problem}

\begin{figure*}[t]
\includegraphics[width=5.4in]{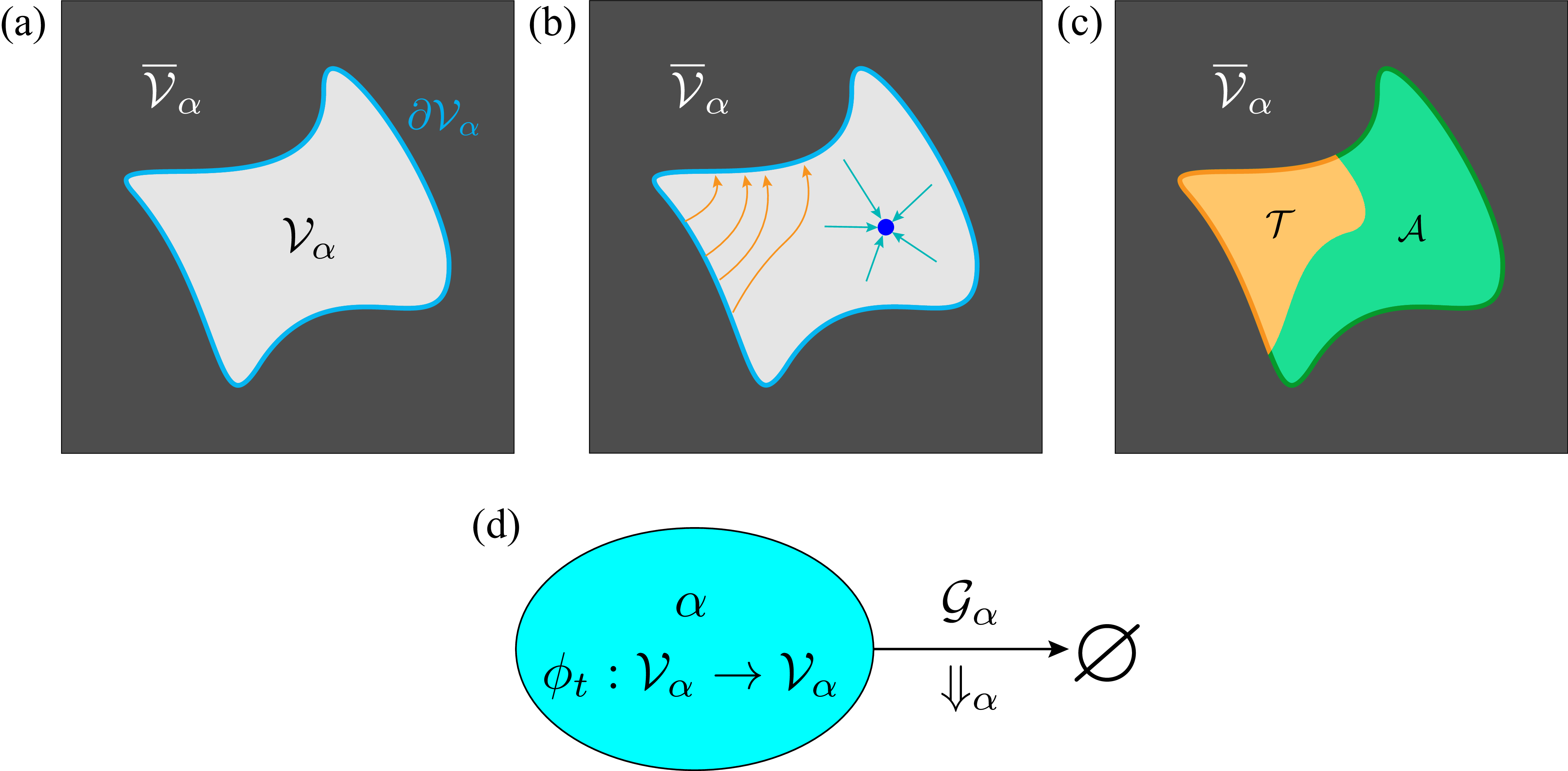}
\caption{\label{fig:epsart} \textbf{(a)} In viability space, a set of constraints creates a viability boundary \(\partial\mathcal{V}_\alpha\) that differentiates the viability region \(\mathcal{V}_\alpha\) where an agent can exist from the complement \(\overline{\mathcal{V}}_\alpha\) where no such life is possible. \textbf{(b)} As the flow \(\phi_t\) unfolds within \(\mathcal{V}_\alpha\), trajectories can reach \(\partial\mathcal{V}_\alpha\) with outward velocity in finite time and die (orange trajectories) or achieve a kind of homeostatic regime as they converge to a viable attractor such as the dark blue equilibrium point (green trajectories). \textbf{(c)} Depending on how the dynamics within the viability region unfold, the space can be decomposed into transiently and asymptotically viable sets, based on whether the initial conditions result in death or persist indefinitely. \textbf{(d)} Formally accounting for death events in our model requires that we transition to a hybrid dynamical system (more specifically, a hybrid automaton), where the discrete modes represented are whether the agent \(\alpha\) is alive or not. The guard conditions \(\mathcal{G}_\alpha\) for a death event are that the agent must be at \(\partial\mathcal{V}_\alpha\) with outward velocity, and the reset function \(\Downarrow_\alpha\) removes the agent's state dimensions from the model.}
\end{figure*}

Work mathematically characterizing the viability of individual agents can be traced back to Ashby \cite{ashbyDesignBrain1960}, who described dynamics as unfolding within a compact region of state space bounded by absolute physiological limits. Today, such limits are commonly referred to as \textit{viability constraints}, and the state variables that they constrain as the agent's \textit{essential variables} \cite{beerDynamicalSystemsPerspective1995}. In this tradition, we will understand an agent to have a set of viability constraints that function as a map \(C_\alpha\), assigning boolean values throughout its state space \(X_\alpha\) according to whether it is alive (note the subscript \(i\) for \(\alpha\) is being dropped since the agent set currently only contains one individual). Importantly, these constraints must be defined to include their limits so that there is a true geometric boundary (i.e. \(\geq\) or \(\leq\) instead of \(>\) or \(<\)): 
\begin{equation}
C_\alpha:X_\alpha\rightarrow\{\mathrm{false},\mathrm{true}\}
\end{equation}
With the agent's viability constraints defined, we can then define its viability region \(\mathcal{V}_\alpha\) as being all the points in the full agent-environment state space \(\mathcal{U}\) where \(C_\alpha\) is true (Fig. 1(a)). Notably, since nonessential variables in the system are not subjected to the viability constraints, the viability region is uniformly extended across these dimensions: 
\begin{equation}
\mathcal{V}_\alpha\triangleq \{\mathbf{x}\in \mathcal{U}:C_\alpha\rm{~is~true} \}
\end{equation}
In the previous subsection, we defined the autonomous dynamics over \(\mathcal{U}\) as being co-determined by the agent and environment, but now we need to account for the fact that the agent's dynamics are only applicable within \(\mathcal{V}_\alpha\). We will do this by assuming the vector field is undefined beyond the viability region, such that the flow \(\phi\) does not cross the boundary of the viability region \(\partial\mathcal{V}_\alpha\):
\begin{equation}
 \dot{\mathbf{x}} = \begin{cases} \mathbf{F}(\mathbf{x}) & \text{if  } \mathbf{x}\in \mathcal{V}_\alpha\\ \mathrm{undefined}& \text{otherwise} \end{cases}
\end{equation}
While this definition makes it clear that the agent does not exist beyond its viability region, we need to go a step further to define its death in the model. This can be accomplished by reformulating the system as a hybrid automaton, combining continuous dynamics (living behavior) with discrete transitions between modes (death) (Fig. 1(d)) \cite{aiharaTheoryHybridDynamical2010}. Importantly, one must not confuse the concepts of \textit{dying} and \textit{death} at this point. Dying is a time-extended process and a part of life, meaning it takes place within the agent's viability region. Death, meanwhile, is the punctuated moment at the end of this process, where the viability constraints are violated, and the agent is irreversibly lost. This perspective emphasizes that death is an event that occurs in the transient activity of a system, long before asymptotics come into play \cite{johLyseNotLyse2011,cortesBenchKeyboardBack2021,mcshaffreyMattersLifeDeath2026}. Even equating thermal equilibrium between an organism and its environment with death is, in some ways, misguided \cite{schrodingerWhatLifePhysical1992}; by the time such an attracting state has been reached, the functioning organism is already long gone. Explicitly representing the transition from life to death requires two features (Fig. 1(d)). The first is a guard condition \(\mathcal{G}_\alpha\) that specifies that the agent only meets the condition for death if it is at the viability boundary with outward velocity, where for \(\nabla\partial\mathcal{V}_\alpha|_{\mathbf{x}}\) we pick the orthogonal vector that points into \(\overline{\mathcal{V}}_\alpha\) as opposed to the interior. Note that if the viability boundary is a piecewise smooth boundary, at sharp junctures there may be more than one vector to evaluate velocity against: 
\begin{equation}
\mathcal{G}_\alpha= \{ \mathbf{x}\in\partial\mathcal{V}_\alpha:\mathbf{F}(\mathbf{x})\cdot\nabla\partial\mathcal{V}_\alpha|_\mathbf{x}>0\}
\end{equation}
The second addition is a reset function that determines what discrete transformation occurs when the guard conditions are met. In this work, the reset takes the form of a \textit{collapse function} \(\mathcal{\Downarrow}_\alpha\). Unlike other hybrid automata, this function does not define a change in the values of state variables, but a removal of the agential dimensions. In the single-agent case, this looks like a map from the viable set of the unified agent-environment system \(\mathcal{V}_\alpha\) down to the environmental state space \(X_{\mathcal{E}}\), restricted to being evaluated where the guard conditions are met:
\begin{equation}
\Downarrow_\alpha|_{\mathcal{G}_\alpha}:\mathcal{V}_\alpha\rightarrow X_\mathcal{E}
\end{equation}
With hybrid automaton fully defined, we can then separate initial conditions in the viability region into two broad classes: \textit{asymptotically} and \textit{transiently} viable (Figs. 1(b) and (c)). Asymptotically viable initial conditions \(\mathcal{A}\) are those where, for all time, the resulting trajectory never satisfies the guard condition for death. Importantly, points at the viability boundary can be asymptotically viable as long as they do not have outward velocity: 
\begin{equation}
\mathcal{A}= \{ \mathbf{x}\in\mathcal{V}_\alpha:\forall t \geq0,\phi_t(\mathbf{x})\notin \mathcal{G}_\alpha\}
\end{equation}
Transiently viable initial conditions \(\mathcal{T}\) are then those that do satisfy the guard conditions at some unique point in time:
\begin{equation}
\mathcal{T}= \{ \mathbf{x}\in\mathcal{V}_\alpha:\exists ! t \geq0,\phi_t(\mathbf{x})\in \mathcal{G}_\alpha\}
\end{equation}
These classifications reveal that agent-environment models with viability constraints form a class of extended dynamical systems where both asymptotic and transient activity must be considered. Further, since initial conditions that would have been in a viable attractor's basin can have their paths cut off by \(\partial\mathcal{V}_\alpha\), fundamentally new organizing manifolds must structure the global behavior of these systems. In the next two sections, we develop \textit{viability space decomposition} as a means to capture these features, culminating in the concept of a \textit{viability portrait}. We begin by developing the theory for single-agents. 

\section{Single-Agent Viability Space Decomposition}

\subsection{Robustness conditions}

\begin{figure*}[p]
\includegraphics[width=\textwidth]{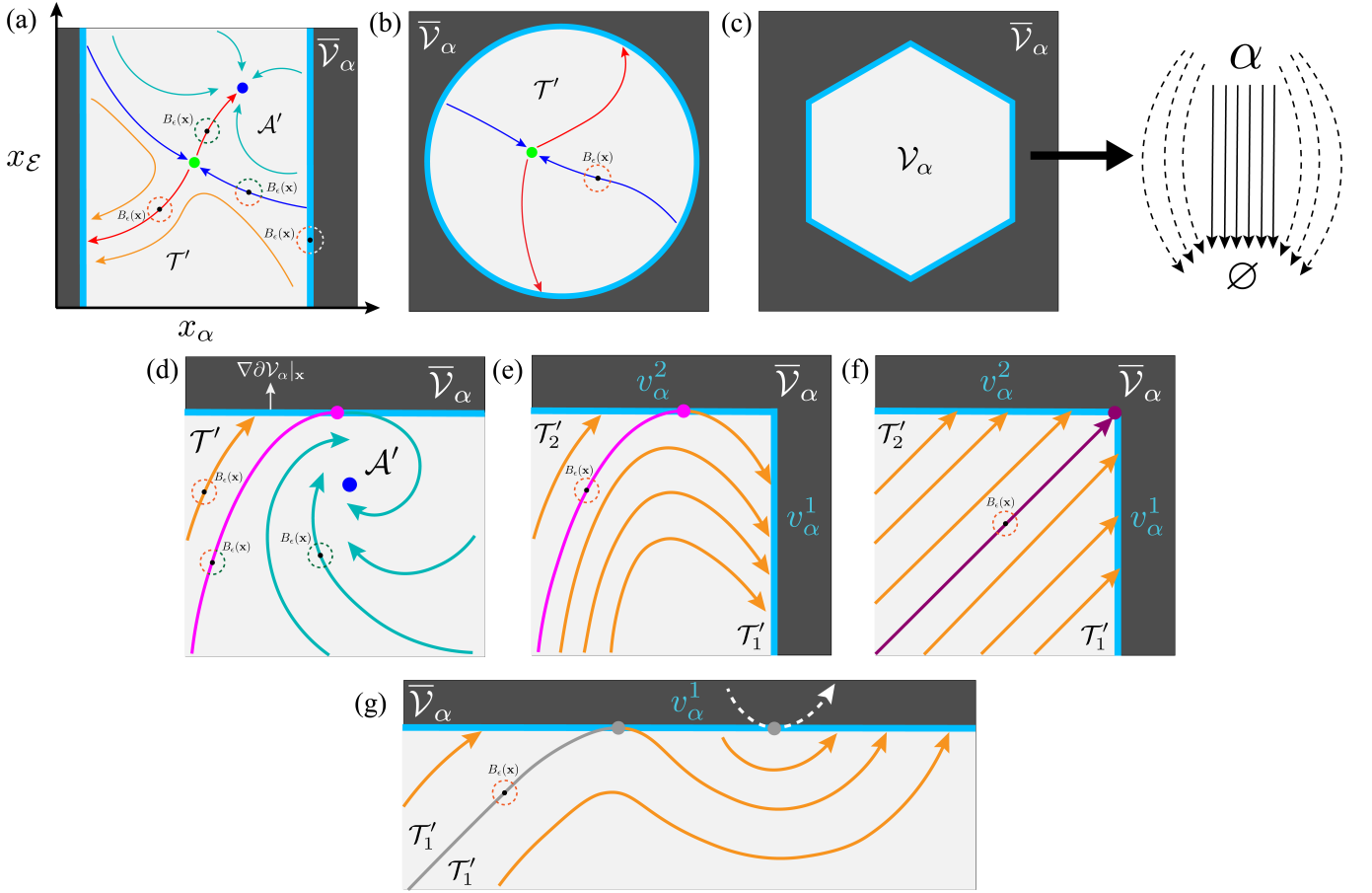}
\caption{\label{fig:epsart} \textbf{(a)} Initial conditions can be categorized as robustly transiently (orange) or asymptotically (green) viable if there exists an open ball \(B_\epsilon\) around that point where all its initial conditions maintain that same fate. Points along \(\partial\mathcal{V}_\alpha\) are not robust by default since any \(B_\epsilon\) overlaps with \(\overline{\mathcal{V}}_\alpha\) where the dynamics are undefined and no agent exists. An attractor in the viability region interior (blue point) indicates a corresponding robust asymptotically viable set, whereas initial conditions with no corresponding attractor will generically be robustly transiently viable. The stable saddle manifold  \(W^s\) (dark blue trajectories) separates the distinct fates reached by the unstable manifold \(W^u\) (red trajectories). While the \(W^s\) trajectories are asymptotically viable since they converge to the saddle-type limit set (light green point), this is not a robust fate outcome because open balls centered on points in \(W^s\) will not all converge to the same fate. \textbf{(b)} If the trajectories that comprise \(W^u\) have the same fate, \(W^s\) will be a \textit{virtual separatrix}, where the regions on either side have qualitatively the same survival outcome. \textbf{(c)} The possible unique transitions that can occur during a death event can be summarized by translating the geometry of \(\partial\mathcal{V}_\alpha\) into an agent-graph, where solid edges correspond to a single smooth segment of the viability boundary, and dashed edges correspond to junctions where multiple segments meet. The dynamics then enable a subset of these edges and prune the rest. \textbf{(d)} Having a viability boundary introduces the possibility that trajectories in a viable attractor's would-be-basin can have their path intercepted, making them transiently viable. We can begin to derive novel boundaries to explain this by looking at how the dynamics interact with \(\partial\mathcal{V}_\alpha\), selecting the orthogonal vector \(\nabla\partial\mathcal{V}_\alpha|_\mathbf{x}\) that points into  \(\overline{\mathcal{V}}_\alpha\). To continuously transition from a portion of \(\partial\mathcal{V}_\alpha\) where the dynamics are fatal to a portion where the dynamics move into the interior, we must pass through an intermediate point where the dynamics are tangent (magenta point). If the tangent point's forward-time trajectory is asymptotically viable, the tangent point is a \textit{mortality point} whose reverse-time trajectory (magenta contour) will be a \textit{mortality manifold} that separates robust transiently viable and robust asymptotically viable trajectories. \textbf{(e)} If we break a piecewise smooth \(\partial\mathcal{V}_\alpha\) into connected smooth segments \(v_\alpha^i\), robust transiently viable initial conditions will be those that have an open ball where all trajectories violate the same \(v_\alpha^i\). Mortality manifolds (magenta) can separate distinct robust transiently viable sets that border each other if the mortality point (magenta) separates terminal vectors on one boundary segment \(v_\alpha^i\) from recovery vectors on the same segment that eventually violate a different segment \(v_\alpha^j\). \textbf{(f)} If robust transiently viable sets border each other, they can also be separated by an \textit{ordering manifold} (purple contour) which can be found by finding the \textit{ordering points} (purple point) where both corresponding guard conditions are satisfied simultaneously and integrating backward in time. \textbf{(g)} If a tangent point (left gray point) separates fatal and recovery vectors along a segment of \(\partial\mathcal{V}_\alpha\) but then returns to violate that same segment, it is not a mortality point and its reverse-time trajectory (gray contour) will not separate distinct survival outcomes. This can happen if, as we continue to move along the viability boundary segment, there is another tangency (right gray point) that marks the transition back to terminal change vectors. This second tangency can be "invisible" in the sense that the curvature indicates the trajectory to either side would be in \(\overline{\mathcal{V}}_\alpha\) where the flow is not defined (white dashed trajectory).
}
\end{figure*}

What additional structure beyond phase portraits must be identified in order to develop a comprehensive picture of survival outcomes in single-agent models? While breaking \(\mathcal{V}_\alpha\) into subsets of asymptotically and transiently viable initial conditions gives us a coarse set of categories, we are usually interested in those initial conditions whose behavior is resilient in the face of imprecise measurement and small perturbations. Therefore, we can further break down these points into robust subsets. In the case of \(\mathcal{A}\), we will assume that an agent achieving a homeostatic regime with its environment requires the presence of some kind of attractor. A point \(\mathbf{x}\) is then part of the robust asymptotically viable set \(\mathcal{A}_i'\) if there exists some nonzero \(\epsilon\)-sized open \(B_\epsilon(\mathbf{x})\) around it such that all points converge to the same viable attractor \(\Omega_i\) (dark blue points in Figs. 2(a) and (c)).\footnote{Here we are denoting attractors with \(\Omega\) because we are not specifying the type of limit set (equilibrium point, limit cycle, etc). When we do, stability class will be denoted with a subscript, using \(\Omega\) for attractors.} This means that there will be a connected, robust asymptotically viable subset for each attractor in the interior of \(\mathcal{V}_\alpha\). While points in \(\partial\mathcal{V}_\alpha\) can be asymptotically or transiently viable, an open ball around them will always overlap with \(\overline{\mathcal{V}}_\alpha\) where the flow is not defined according to Eqn. 5, meaning these initial conditions are never robust in their viability outcome (Fig. 2(a)). Relatedly, if a would-be attracting limit set has tangent contact with \(\partial\mathcal{V}_\alpha\), it is not a true attractor, resulting in a degenerate configuration within the formalism, and any would-be limit set that extends into \(\overline{\mathcal{V}}_\alpha\) no longer qualifies for having lost its invariance:
\begin{equation}
\mathcal{A}_i'=\{\mathbf{x}\in\mathcal{A}:\exists\epsilon>0, \forall \mathbf{y}\in B_\epsilon(\mathbf{x}),\phi_{t\rightarrow\infty}(\mathbf{y})\in\Omega_i \}
\end{equation}
Conversely, the robust transiently viable initial conditions \(\mathcal{T}'\) are those where there exists some nonzero \(\epsilon\)-sized open ball such that all the points in it meet the guard condition for death. This robust set may then be further decomposed into disconnected subsets separated by asymptotically viable initial conditions:
\begin{equation}
\mathcal{T}'=\{\mathbf{x}\in\mathcal{T}:\exists\epsilon>0, \forall \mathbf{y}\in B_\epsilon(\mathbf{x}),\mathbf{y}\in\mathcal{T}\}
\end{equation}
Similar to how we identified robustly asymptotically viable subsets by their capacity to remain within \(\mathcal{V}_\alpha\) as well as which viable attractor they converge to, we have the option to more finely decompose \(\mathcal{T}\) based on whether the initial conditions robustly result in the same cause of death, as opposed to death in general. Assuming that \(\partial\mathcal{V}_\alpha\) is a piecewise-smooth boundary, and every smoothly varying segment corresponds to the boundary of some unique constraint in \(C_\alpha\), the most intuitive option is to view the boundary as being the union of these segments: 
\begin{equation}
\partial\mathcal{V}_\alpha=\bigcup_{i=1}^k v^i_\alpha 
\end{equation}
This allows us to accordingly break up the guard condition \(\mathcal{G}_\alpha\) based on which segment of \(\partial\mathcal{V}_\alpha\) the dynamics are violating:
\begin{equation}
\mathcal{G}_\alpha=\bigcup_{i=1}^k\mathcal{G}^i_\alpha 
\end{equation}
Importantly, where multiple segments meet at sharp junctures, it is possible for the dynamics to violate multiple constraints simultaneously if all the corresponding guard conditions are met. As a rule of thumb, death outcomes are more likely to occur when the dimensionality of the corresponding viability segments is not significantly different than the dimensionality of the viability region. For example, if \(\mathcal{V}_\alpha\) is an \(n\)-dimensional space, the segments of its viability boundary will be \(n-1\)-dimensional, the intersection of two segments will be \(n-2\)-dimensional, and so forth. This relationship can be compactly represented by a measure of codimension, where \(\{v^i_\alpha\cap\ldots\}\) is the set of one or more boundary segments, and \(\mathrm{dim}(\ldots)\) is a measure of dimensionality:
\begin{equation}
\mathrm{codim}(\{v^i_\alpha\cap\ldots\})=\mathrm{dim}(\mathcal{V}_\alpha)-\mathrm{dim}(\{v^i_\alpha\cap\ldots\})
\end{equation}
Measuring codimension reveals an important detail about the geometry of death outcomes in the interior of the viability region. If every smoothly varying segment \(v^i_\alpha\) of \(\partial\mathcal{V}_\alpha\) is a codimension 1 structure, the family of trajectories that violates it will generally be codimension 0. In other words, it will be a (hyper)volume of equal dimension to \(\mathcal{V}_\alpha\). Similar to how robust asymptotically viable points have an \(\epsilon\)-sized open ball where all initial conditions still converge to the same attractor, this means we can expect to find an \(\epsilon\)-sized open ball around those transiently viable initial conditions \(\mathcal{T}_i'\) that die from violating a unique segment of the viability boundary, excluding boundary points themselves. Notably, it is still true that the robust set may contain multiple disconnected subsets if they are separated by asymptotically viable initial conditions:
\begin{eqnarray}
\mathcal{T}_i'=\{\mathbf{x}\in\mathcal{T}:\exists! v^i_\alpha,\exists\epsilon>0, \forall \mathbf{y}\in B_\epsilon(\mathbf{x}),\nonumber \\
 \exists ! t\geq0,\phi_t(\mathbf{y})\in\mathcal{G}^i_\alpha\}
\end{eqnarray}
Any families of trajectories that violate two or more segments of the viability boundary simultaneously will only form a geometry of codimension 1 or higher (purple trajectory in Fig. 2(f)). This means that, while these death events can certainly occur, there will not be an open ball that uniquely contains them, and therefore they cannot be characterized as robust. Nonetheless, it is still important to account for all the possible death outcomes in a model. This can be compactly represented by forming an \textit{agent-graph} (Fig. 2(c)), where there are two nodes, representing the states of the world when the agent is alive and where it is not. There is then a solid directed edge for each segment of the viability boundary, representing the possibility of a unique cause of death, and a dashed directed edge for each possible transition that involves two or more simultaneous causes of death. This collective set of edges can then be pruned based on which transitions are enabled by the dynamics. While all these transitions result in the loss of the agent, there are many cases in biology where we want to know which specific death event occurred. For example, regulated cell death processes such as apoptosis unfold in a far more controlled manner than unregulated processes such as necrosis, and therefore have very different implications for multicellular physiology \cite{yuanGuideCellDeath2024}. 

\subsection{Saddle manifolds and fate outcomes}

Given these definitions for robustly observable survival outcomes in the viability region, is it possible to identify global manifolds that organize the entire space, analogous to how we'd decompose state space in a classical dynamical systems analysis? The intuitive first step is to see how far traditional separatrices alone can get us. In one-dimensional dynamical systems, identifying the stable and unstable equilibria allows for a complete analysis, with the unstable equilibria functioning as the boundaries between any two robust regions of existential outcomes. Similarly, the stable manifolds \(W^s\) of saddle-type objects can separate robust sets in higher-dimensional dynamical systems depending on the survival outcomes of trajectories in the unstable manifold \(W^u\). For example, in Fig. 2(a), there is one viable attractor with a corresponding asymptotically viable set, and one robust transiently viable set. Since one branch of \(W^u\) converges to the attractor and the other violates a segment of the viability boundary, \(W^s\) will separate the \(\mathcal{T}'\) and \(\mathcal{A}'\) sets these trajectories belong to. While both branches of \(W^s\) are asymptotically viable, they are not robustly so since the saddle node is not an attractor, and points within open balls centered along the branches will lead to different survival outcomes. Alternatively, if the trajectories that make up \(W^u\) all violate the same continuous constraint, \(W^s\) exists but does not separate distinct types of death outcomes (Fig. 2(b)). In this context \(W^s\) forms a virtual separatrix, analogous to what is seen in several classical dynamical systems \cite{abrahamDynamicstheGeometryBehavior1992}. 

While saddle manifolds can still form meaningful boundaries in viability space, in two and higher dimensions they are often not sufficient for a complete decomposition. For example, stable saddle manifolds traditionally form the basin boundaries for attractors, but now there is the possibility that viability constraints will cut off trajectories that would have otherwise converged to such a stable state. Furthermore, saddle manifolds may not always distinguish between neighboring transiently viable sets that violate distinct segments of the viability boundary. Both of these complications demand the existence of novel global structures. 

\subsection{Mortality manifolds}

In order to understand how survival outcomes are organized throughout the interior of \(\mathcal{V}_\alpha\), we need to start by more carefully considering how the dynamics interact with \(\partial\mathcal{V}_\alpha\), the boundary between life and death. If we look along \(\partial\mathcal{V}_\alpha\), the two most common change vectors we can expect to see are those that move the state into the interior of \(\mathcal{V}_\alpha\) (recovery) and those that have outward velocity (fatality) (Figs. 2(d), (e), and (g)). Along any smoothly varying boundary segment \(v_\alpha^i\), continuously transitioning from recovery to fatality vectors requires that we pass through a point where the change vector is exactly tangent to \(\partial\mathcal{V}_\alpha\). The set of tangent points \(\Lambda\) therefore functionally separates the boundary points that recover from those that result in instantaneous death:
\begin{equation}
\Lambda= \{ \mathbf{x}\in\partial\mathcal{V}_\alpha:\mathbf{F}(\mathbf{x})\cdot\nabla\partial\mathcal{V}_\alpha|_{\mathbf{x}}=0\}
\end{equation}
While having an organizing set for instantaneous outcomes at the viability boundary is helpful, it does not clarify what happens past this first moment. The first opportunity to organize global dynamics appears if we restrict \(\Lambda\) to those points that also remain asymptotically viable, meaning that the tangency marks a transition from transiently viable initial conditions in \(\partial\mathcal{V}_\alpha\) to those that converge to a viable attractor. Such tangencies are an instance of a special class of organizing point which we will call \textit{mortality points} \(\mathbf{m}\) (magenta point in Fig. 2(d)):
\begin{equation}
\mathrm{if~}\mathbf{x}\in\bigcap\{\mathcal{A},\Lambda\},\mathrm{then~}\mathbf{x}\mathrm{~is~a~mortality~point}
\end{equation}
These points \(\mathbf{m}\), while leading to robustly asymptotically viable trajectories, are not themselves robust since an open ball around them would contain fatal vectors, the transiently viable points that lead to them, other points in \(\partial\mathcal{V}_\alpha\), and points in \(\overline{\mathcal{V}}_\alpha\). Nonetheless, these points are still important to identify, as their \textit{reverse-time} trajectories form a \textit{mortality manifold} \(\mathcal{M}\) that is a boundary between the robust transiently viable initial conditions that will violate a segment of the viability boundary and the robust asymptotically viable points that will never make contact with \(\partial\mathcal{V}_\alpha\) (Fig. 2(d)). In two-dimensional dynamical systems, integrating a single mortality point backward in time gives a mortality manifold \(\mathcal{M}\), noting that Eqn. 5 means that the flow will not be able to carry the trajectory past where the dynamics are undefined:
\begin{equation}
\mathcal{M}=\{\phi_t(\mathbf{m}):t\leq0\}
\end{equation}
In higher dimensional systems, a family of mortality points \(\mathbf{M}\) will be required to form \(\mathcal{M}\). In general, families of mortality points will be codimension 2 sets of \(\mathcal{V}_\alpha\) embedded in \(\partial\mathcal{V}_\alpha\). When integrated backward in time, these families extend to form codimension 1 organizing boundaries that separate \(\mathcal{T}'\) and \(\mathcal{A}'\) sets: 
\begin{equation}
\mathcal{M}=\bigcup_{i\in \mathbf{M}}\{\phi_t(\mathbf{m}_i):t\leq0\}
\end{equation}
If we are only interested in distinguishing the fates of initial conditions based on whether they result in survival or death, then traditional phase portrait structures and \(\mathcal{M}\) are sufficient, and Eqn. 17 can be strengthened to a bi-conditional. But, if we are also interested in distinguishing between types of deaths corresponding to violating the different segments of  \(\partial\mathcal{V}_\alpha\), additional machinery is still needed. First of all, the concept of mortality manifolds can be generalized. As currently presented, mortality manifolds separate robust transiently and asymptotically viable sets because mortality points are as close as the dynamics can get to death while still leading to asymptotically viable trajectories. In a similar manner, mortality points can also separate bordering transiently viable sets if they lie tangent to one boundary segment \(v_\alpha^i\) and proceed to meet the guard condition \(\mathcal{G}^j_\alpha\) of another segment \(v_\alpha^j\) (Fig. 2(e)). From there, Eqns. 18 and 19 still apply: 
\begin{eqnarray}
\mathrm{if~}\mathbf{x}\in\Lambda\cap v_\alpha^i\wedge\exists ! t\geq0:\phi_t(\mathbf{x})\in\mathcal{G}^j_\alpha ,\nonumber \\ \mathrm{then~}\mathbf{x}\mathrm{~is~a~mortality~point}
\end{eqnarray}
If the trajectory from the tangency meets the guard condition for its originating boundary segment, it is not a mortality point and the reverse-time trajectory will not globally organize distinct fate outcomes (Fig. 2(g)), similar to how the stable saddle manifold does not always separate distinct outcomes if the unstable manifold trajectories all lead to the same fate (Fig. 2(b)). 

\subsection{Ordering manifolds}

Another way neighboring robust transiently viable sets can be distinguished is if there are initial conditions that lead to multiple death outcomes happening simultaneously. More specifically, we start by looking at the intersections of the viability boundary segments and see if there are \textit{ordering points} \(\mathbf{o}\) where all corresponding guard conditions are satisfied: 
\begin{equation}
\mathbf{x}\mathrm{~is~an~ordering~point}\Leftrightarrow\mathbf{x}\in\bigcap\{\mathcal{G}^i_\alpha,\mathcal{G}^j_\alpha,\ldots\}
\end{equation}
In two dimensional dynamical systems, a single ordering point where two guard conditions are met simultaneously can be integrated backward in time to form an \textit{ordering manifold} \(\mathcal{O}\) that separates two robust transiently viable regions (purple trajectory in Fig. 2(f)):
\begin{equation}
\mathcal{O}=\{\phi_t(\mathbf{o}):t\leq0\}
\end{equation}
Like the mortality manifolds, ordering manifolds in higher dimensional dynamical systems will originate from the reverse-time trajectories of a family of ordering points \(\mathbf{O}\). Since ordering manifolds originate from the sharp junctures where multiple segments of the viability boundary meet, the backward time extension may be a stratified manifold with piecewise smooth elements, similar to the viability boundary itself:
\begin{equation}
\mathcal{O}=\bigcup_{i\in \mathbf{O}}\{\phi_t(\mathbf{o}_i):t\leq0\}
\end{equation}
This culmination of global structures presented in Fig. 2 provides a complete skeleton of the possible existential outcomes in the single-agent model: a \textit{viability portrait}. Next we extend this framework to the multi-agent case. 

\section{Multi-Agent Viability Space Decomposition}
\begin{figure*}[t]
\includegraphics[width=\textwidth]{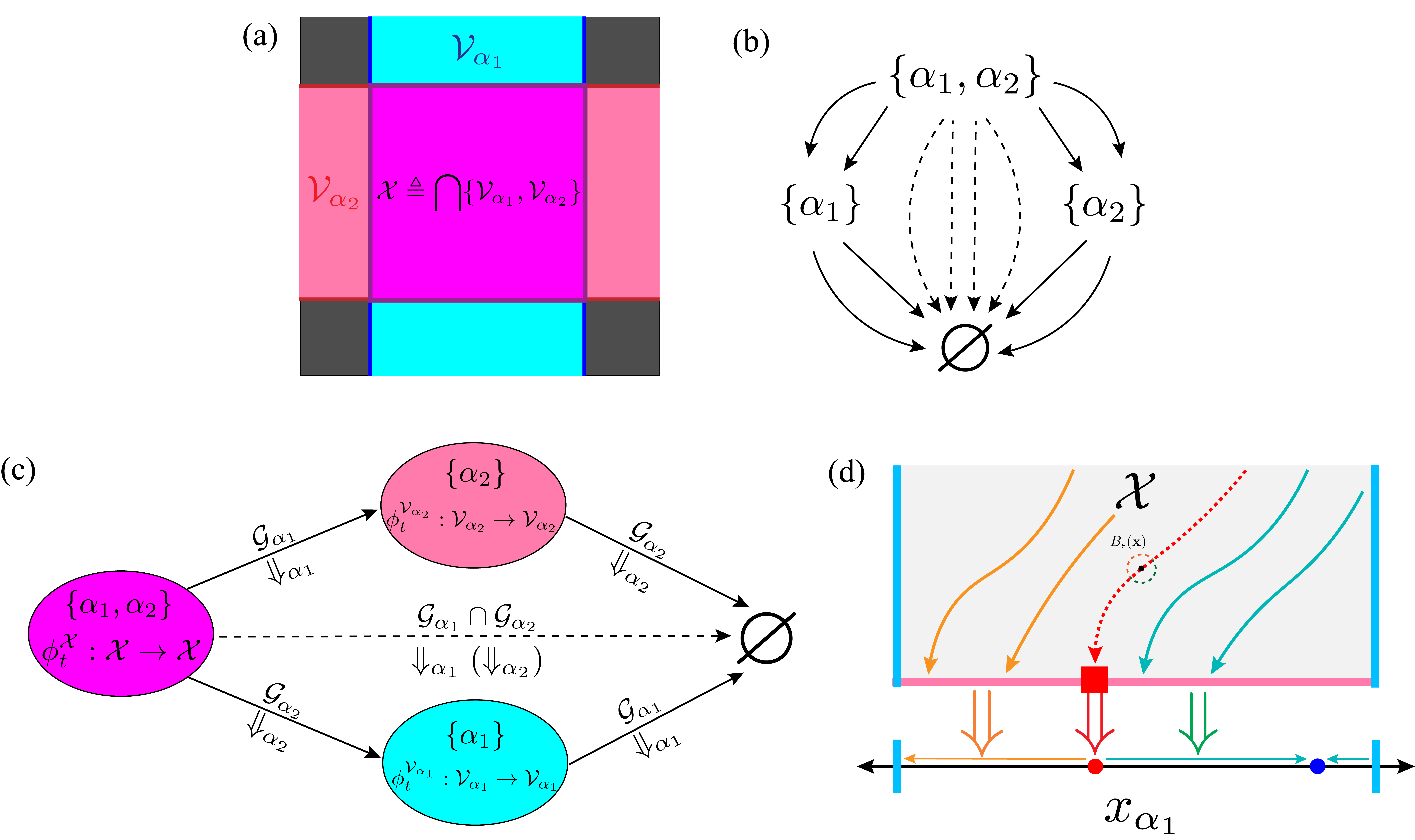}
\caption{\label{fig:epsart} \textbf{(a)} The domain of interaction of the agents is defined as the intersection of their viability regions, denoted \(\mathcal{X}\). \textbf{(b)} Depending on how the viability constraints are entangled, the death possibilities can be represented as a directed graph whose vertices are the power set of the agent list. The graph edges correspond to the constraints in Fig. (a). There is a solid edge for each constraint that could result in an agent's death, and the dashed edges correspond to higher codimension portions of \(\partial\mathcal{X}\) where multiple constraints meet. If a portion of \(\partial\mathcal{X}\) belongs to more than one agent, simultaneous death events become possible. \textbf{(c)} Accounting for the dynamics and deaths of multiple agents requires extending the hybrid automaton. For each configuration of living agents, there is a unique flow for that domain. Each agent has its own guard conditions for death and a collapse function that removes its state variables if those conditions are satisfied. A transition that results in simultaneous deaths (dashed edge) must occur at the intersection of guard conditions and reset through a composition of the corresponding collapse functions. \textbf{(d)} Analyzing a multi-agent system requires decomposing sets of initial conditions not only based on the existential outcomes in the current domain, but also on the fates of the agents that survive the first death event. Here, we see that in \(\mathcal{X}\), all of the trajectories result in the death of \(\alpha_2\), but depending on the state \(\alpha_1\) is in when this occurs, the collapse function may leave it in a transiently or asymptotically viable portion of its state space. These two regions in \(x_{\alpha_1}\) are separated by an unstable equilibrium point (red dot). By mapping this unstable point onto the death event that collapses exactly onto it (red square), we can then integrate this point backward in time to get a \textit{collapse manifold} (red dashed contour), which further decomposes \(\mathcal{X}\) into robust sets that account for the fates of all agents in the agent list. An open ball \(B_\epsilon\) centered on the collapse manifold contains points that differ in fate for \(\alpha_1\) even though they all lead to the same fate for \(\alpha_2\). Any organizing manifold that occurs in a lower-dimensional domain of the hybrid automaton has the potential to be extended into connected higher-dimensional domains as a collapse manifold.}
\end{figure*}

\subsection{Domains of interaction}

Generalizing the viability space decomposition to models that have any number of interacting agents \(\alpha=\{\alpha_1,\ldots,\alpha_N\}\) requires that we shift our attention from any single individual's viability constraints to how the collective set of constraints forms a range of states that all agents can exist and interact in. We will refer to this range of states as a \textit{domain of interaction} \(\mathcal{X}\) \cite{varelaPrinciplesBiologicalAutonomy2025, maturanaAutopoiesisCognitionRealization1980}, defined as the intersection of all agents' viability regions: 
\begin{equation}
\mathcal{X}\triangleq\bigcap_{i=1}^N \mathcal{V}_{\alpha_i}
\end{equation}
There is a unique domain of interaction for every combination of agents in the world, and depending on which agents die and in which order, the overall system may end up in any of these domains. This requires that we extend the hybrid automaton to have a number of nodes \(\mathcal{N}\) equal to the cardinality of the power set of the agent-list \(\alpha\), where the cardinality of the agent list \(|\alpha|\) corresponds to the number of nodes where a single agent is alive, and one node will correspond to where the agent set is empty:
\begin{equation}
\mathcal{N}=|\mathcal{P}(\alpha)|=2^{|\alpha|}
\end{equation}
Each of these domains of interaction will have a unique flow collectively shaped by the agents currently alive and the environment:
\begin{equation}
\phi_t^\mathcal{X}:\mathcal{X}\rightarrow\mathcal{X}
\end{equation}
And like in Eqns. 6, 12, and 13, each agent will have guard conditions \(\mathcal{G}_{\alpha_i}\) that establish its death conditions, and these guard conditions can be further broken up so they correspond to the piecewise smooth segments of viability boundaries:
\begin{equation}
\partial\mathcal{V}_{\alpha_i}=\bigcup_{j=1}^k v^j_{\alpha_i} 
\end{equation}
\begin{equation}
\mathcal{G}_{\alpha_i}=\bigcup_{j=1}^k \mathcal{G}^j_{\alpha_i} 
\end{equation}
The presence of more than one agent has important implications for the variety of discrete transitions that can be made in the hybrid automaton. Like in the single-agent case, if an agent's viability boundary segments are violated, a collapse function is activated which removes that agent from the model by getting rid of its corresponding state dimensions:
\begin{equation}
\Downarrow_{\alpha_i}|_{\mathcal{G}_{\alpha_i}}:\partial\mathcal{X}_{\mathrm{current}}\rightarrow\mathcal{X}_{\mathrm{new}}
\end{equation}
Alternatively, if simultaneously triggered guard conditions belong to more than one agent, what we actually have is a rare event where multiple agents are lost, requiring us to compose collapse functions:
\begin{equation}
\Downarrow_{\alpha_i,\ldots,\alpha_j}=\Downarrow_{\alpha_i}\circ\ldots\circ\Downarrow_{\alpha_j}
\end{equation}
This means that the set of directed edges in the hybrid automaton depends on how the viability boundary segments of agents are entangled in \(\partial\mathcal{X}\). As before, single viability boundary segments are codimension 1 structures, and trajectories that violate them will generally exist in a codimension 0 (hyper)volume of state space. As in Eqn. 14, the intersection of two or more boundary segments is a higher codimension structure (Fig. 3(a)). The main complication introduced now is that if the guard conditions met belong to more than one agent, multiple agents can die simultaneously, which impacts the domain of interaction that the composition of collapse functions will bring the system to. As before, these death possibilities based on the geometry generated by constraints can be compactly represented in an agent-graph with \(\mathcal{N}\) nodes (Eqn. 25), solid directed edges for each co-dimension 1 boundary segment, and dashed edges for geometries of higher codimension where multiple death conditions can be simultaneously satisfied (Fig. 3(b)). 

How are the existential outcomes organized in these more complex multi-agent systems? As before, the key idea is to identify robust subsets where the qualitative dynamics are not disrupted by noise and imprecision, and then derive the manifolds that separate them. As in the single-agent case, continuous dynamics within the hybrid automaton can either remain in \(\mathcal{X}\) by reaching a viable attractor, or violate \(\partial\mathcal{X}\) with outward velocity. Notably, the attractors that allow an agent to maintain homeostasis are no longer just a byproduct of its relationship with the environment, but its reciprocal interaction with other agents. To denote that an attractor exists in a particular domain of interaction \(\mathcal{X}\), we will include a superscript \(\Omega^\mathcal{X}_i\). Correspondingly, we will drop the prime notation for the robust asymptotically viable set, and instead give it a superscript for the domain:
\begin{equation}
\mathcal{A}^\mathcal{X}_i
\end{equation}
We will include a similar superscript for robust transiently viable sets within a particular domain of interaction, and make the subscript correspond to the agent that has died: 
\begin{equation}
\mathcal{T}^\mathcal{X}_{\alpha_i}
\end{equation}
For any given domain of interaction, the same analyses from the single-agent theory can decompose the space into robust asymptotically and transiently viable initial conditions, noting that ordering manifolds become more important, as they determine which agent dies when transiently viable regions border each other. But this is no longer sufficient to decompose the entire model. Now, we need to understand not only how an initial condition leads to particular fates within the immediate domain of interaction, but also the fates of the agents that outlive this first fatality, and the order in which they occur. Ultimately, this depends on where the collapse functions land the surviving agents' states relative to the organizing manifolds of the new domain. 

\subsection{Collapse manifolds}

As with stable saddle, mortality, and ordering manifolds, which are found by identifying organizing points and integrating backwards in time, we can identify which death events lead to particular states by looking at the inverse of the collapse function. Since the collapse function \(\Downarrow_{\alpha_i}\) only activates where the agent \(\alpha_i\)'s guard condition \(\mathcal{G}_{\alpha_i}\) is satisfied, its inverse \(\Downarrow_{\alpha_i}^{-1}\) maps from the current domain without \(\alpha_i\) to the points along the previous domain's boundary that kill that agent:
\begin{equation}
\Downarrow_{\alpha_i}^{-1}:\mathcal{X}_{\mathrm{current}}\rightarrow\partial\mathcal{X}_{\mathrm{previous}}\cap\mathcal{G}_{\alpha_i}
\end{equation}
This means that if we have an organizing manifold in a lower-dimensional domain, we can use the inverse of the collapse function to map which terminal points on the boundary of the previous domain collapse exactly onto that manifold. For example, assume we have an unstable equilibrium point \(\mathrm{EP}_\mathrm{u}^{\mathcal{V}_{\alpha_1}}\)  as in Fig. 3(d) (red point), which forms a boundary between two sets of robust outcomes in \(\mathcal{V}_{\alpha_1}\). In order for \(\alpha_1\) to be on its own, \(\alpha_2\) must have perished in their domain of interaction \(\mathcal{X}\). Mapping back to the point(s) that fall directly onto the unstable equilibrium then gives a codimension 2 embedding in \(\partial\mathcal{X}\) (red square):
\begin{equation}
\Downarrow_{\alpha_2}^{-1}(\mathrm{EP}_\mathrm{u}^{\mathcal{V}_{\alpha_1}})=\tilde{\mathrm{EP}}_\mathrm{u}^{\mathcal{V}_{\alpha_1}}\in\mathcal{G}_{\alpha_2}
\end{equation}
This unique subset of the fatal vectors can then be integrated backward in time according to the flow of that domain, extending the equilibrium point as a \textit{collapse manifold} \(\mathcal{C}\), where the subscript is the originating manifold and the superscript shows the domain in which the extension exists, followed by the domain of origination: 
\begin{equation}
\mathcal{C}_{\mathrm{EP}_\mathrm{u}^{\mathcal{V}_{\alpha_1}}}^{\mathcal{X},\mathcal{V}_{\alpha_1}}=\{\phi_{t}^\mathcal{X}(\tilde{\mathrm{EP}}_\mathrm{u}^{\mathcal{V}_{\alpha_1}}): t \leq0 \}
\end{equation}
Collapse manifolds can be repeatedly extended into new domains for as long as there is an inverse collapse function that maps them to a nonempty terminal set in the boundary of a higher-dimensional domain of interaction. It is also possible to view the entire set of a manifold's inverse collapse extensions as a singular object. For example, we could extend \(\mathcal{C}_\mathrm{EP_u}^{\mathcal{X},\mathcal{V}_{\alpha_1}}\) into a higher-dimensional domain of interaction \(\mathcal{X}_2\) using the inverse of a third agent \(\alpha_3\)'s collapse function \(\Downarrow_{\alpha_3}^{-1}\):
\begin{equation}
\mathcal{C}_{\mathrm{EP}_\mathrm{u}^{\mathcal{V}_{\alpha_1}}}=\mathrm{EP}_\mathrm{u}^{\mathcal{V}_{\alpha_1}}\cup\mathcal{C}_{\mathrm{EP}_\mathrm{u}^{\mathcal{V}_{\alpha_1}}}^{\mathcal{X},\mathcal{V}_{\alpha_1}}\cup\mathcal{C}_{\mathrm{EP}_\mathrm{u}^{\mathcal{V}_{\alpha_1}}}^{\mathcal{X}_2,\mathcal{X}}\cup\ldots
\end{equation}
Extending all of the organizing manifolds allows a complete decomposition of the hybrid automaton, where robust regions in every domain do not just correspond to the fates that will immediately unfold, but also the connected fates the collapse functions lead to in subsequent domains. An \(\epsilon\)-sized open ball around any segment of a collapse manifold's extension will contain initial conditions that \textit{eventually} diverge in their fate outcome, whereas points to either side of this structure will generally be robust in this outcome. Reaching an asymptotically viable region in any domain will stabilize the cascade of collapses. For example, if we have a three-agent system, a robust region where \(\alpha_2\) and \(\alpha_3\) die followed by \(\alpha_1\) reaching an attractor \(\Omega_i^{\mathcal{V}_{\alpha_1}}\) in its viability region will be labeled: 
\begin{equation}
\mathcal{T}^{\mathcal{X}_2}_{\alpha_2}\mathcal{T}^{\mathcal{X}_1}_{\alpha_3}\mathcal{A}^{\mathcal{\mathcal{V}}_{\alpha_1}}_i
\end{equation}
With this, we have a complete theory to explain the global organization of survival outcomes in multi-agent models. Next we turn to specifying the models that we will use to demonstrate the theory. 

\section{Demonstrations in Models}

In this section, we demonstrate viability space decomposition by presenting a complete analysis of three models: a single cell's physiological dynamics, the single cell behaving, and two cells interacting. While viability space decomposition can be applied to any ODE model of agents with life-death boundaries, we focus on cellular systems simply because they are the fundamental organismal unit. For each system, we begin with a numerical exploration of each model, sampling a high-resolution distribution of initial conditions and tracking their fates. With this establishing a ground truth to compare against, we then find the manifolds that globally organize the survival outcomes in each model, culminating in their viability portraits. These maps both explain the results of the numerical exploration and in some cases even reveal fine details that were missed, demonstrating the effectiveness of the theory. 

The model systems presented in this section are not meant to capture the specifics of any actual biological system, but to demonstrate the theory and build our intuitions in dimensions low enough to visualize \cite{beerAnimalsAnimatsWhy2009a,beerMilkingSphericalCow2024a}. That said, each model introduces an additional layer of complexity representative of real problems in biology, illustrating how the framework handles such features. First, the single-cell model represents the fact that organisms have multiple and often interdependent physiological needs, in this case taking the form of an autocatalytic network with a mathematical structure similar to many gene regulatory and enzyme kinetic models \cite{kauffmanInvestigations2000,alonIntroductionSystemsBiology2019,mcshaffreyDecomposingViabilitySpace2023}. The next model complicates this picture by introducing behavior in the form of chemotaxis in an environment with two opposing food gradients, forcing us to characterize the organizing manifolds in a higher-dimensional space \cite{kalininResponsesEscherichiaColi2010}. Finally, the third model considers how interactions between agents can shape fate outcomes in a two-cell system \cite{mcshaffreyDissectingViabilityMultiAgent2024}, loosely inspired by models linking multicellular interaction to tissue patterning \cite{collierPatternFormationLateral1996}. 

The single and behaving cell models are simplified versions of an earlier model that was designed to highlight the complexities of assessing viability when an organism has multiple interdependent needs \cite{mcshaffreyMaintainingViabilityMultiple2022}. Preliminary decompositions were previously presented for both the single cell circuit without behavior \cite{mcshaffreyDecomposingViabilitySpace2023} and the two-cell model \cite{mcshaffreyDissectingViabilityMultiAgent2024}, although neither with the complete framework. All models are presented in a chemical and enzyme kinetic formalism assuming mass action, with square brackets representing concentrations. Some parameters are reused across models, but appropriate values can be found in Table 1.

\begin{table}[b]
\caption{\label{tab:table}%
Parameter values for models.
}
\begin{ruledtabular}
\begin{tabular}{lccc}
\textrm{Symbol}&
\textrm{Single Cell}&
\textrm{Behaving Cell}&
\textrm{Coupled Cells}\\
\colrule
\(r_\mathrm{chemical}\) & 0.04 & 0.04 & N/A \\
\(K_1\) & 4 & 4 & 4 \\
\(h_1\) & 3 & 3 & 3 \\
\(k_d\) & 0.05 & 0.05 & 0.05 \\
\([F_1]\) & 28 & N/A & N/A \\
\([F_2]\) & 6 & N/A & N/A \\
\(r_\mathrm{spatial}\) & N/A & 2 & N/A \\
\(K_2\) & N/A & 0.5 & 1 \\
\(h_2\) & N/A & 20 & 4 \\
\(r_{M_1}\) & N/A & N/A & 0.064 \\
\(r_{M_2}\) & N/A & N/A & 0.048 \\
\([F]\) & N/A & N/A & 10 \\
\(d_1\) & N/A & N/A & 2 \\
\(d_2\) & N/A & N/A & 5 \\
\end{tabular}
\end{ruledtabular}
\end{table}

\subsection{Single-cell physiology}

\subsubsection{Model specification}

The single-cell physiology is composed of two molecular species \(M=\{M_1,M_2\}\) which reciprocally catalyze each other and are synthesized from one of two food molecules \(F=\{F_1, F_2\}\) respectively. This makes the set of molecules autocatalytic. Fig. 4(a) depicts this circuit with normal arrows representing material transformation, rounded-tip arrows representing catalysis, and \(\varnothing\) designating inconsequential waste products. We treat the concentration of food molecules in the environment as a constant, under the assumptions that it is continuously replenished and that the cell uses up a comparatively insignificant amount at any given moment.

Translating the Fig. 4(a) network diagram into differential equations requires defining functions for synthesis and decay. The rate of synthesis for molecule \(M_i\) is determined by the function \(S\), which takes the other catalyzing molecule \(M_j\) as an input. \([M_j]\) changes this rate through a standard excitatory Hill function, which causes saturation of the rate at high and low concentrations. Here \(K_1\) is the concentration needed for half-maximal activation of the Hill function, and \(h_1\) establishes the steepness of this curve. The parameter \(r_{\mathrm{chemical}}\) sets the maximum rate for the Hill function, and this is multiplied by the available concentration of the corresponding food molecule \([F_i]\):
\begin{equation}
S([M_j]) = r_{\mathrm{chemical}} \left( \frac{[M_{j}]^{h_1}}{K_1^{h_1}+[M_j]^{h_1}} \right) [F_i]
\end{equation}
The decay of each molecule is its current concentration \([M_i]\) multiplied by rate constant \(k_d\):
\begin{equation}
D([M_i]) = k_d[M_i]
\end{equation}
Combining the functions \(S([M_j])\) and \(D([M_i])\) for each molecule gives us the coupled differential equations for the cell's physiology:
\begin{equation}
\frac{d[M_1]}{dt} = S([M_2])-D([M_1])
\end{equation}
\begin{equation}
\frac{d[M_2]}{dt} = S([M_1])-D([M_2])
\end{equation}
Both of these molecular concentrations will play the role of essential variables, factoring into a total of three viability constraints. The first two constraints establish that if either molecular concentration drops below \(0.1\), the cell will perish. The third constraint is a function of the sum of concentrations in the cell, and assumes that if this value exceeds \(20\) then the cell will have an osmotic crisis and burst. According to Eqns. 4 and 5, this then defines the bounded range of states where the differential equations are applicable: 
\begin{equation}
\mathcal{V}_{\alpha}\triangleq \Bigl\{ \mathbf{x}\in \mathcal{U}:\forall[M_i]\geq0.1 \wedge \sum_i [M_i]\leq20 \Bigl\}
\end{equation}
\begin{figure}[t]
\includegraphics[width=3in]{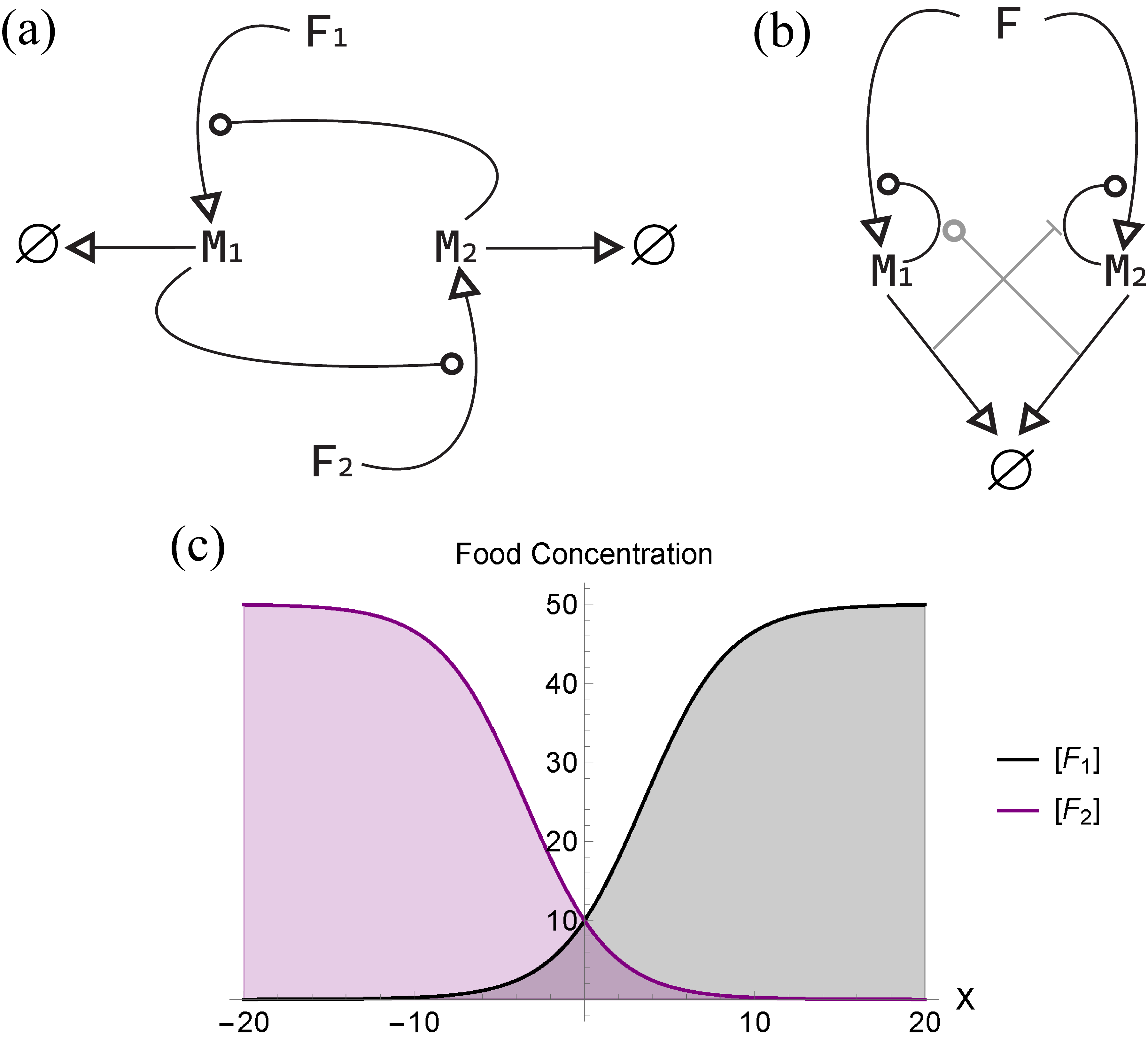}
\caption{\label{fig:epsart} In the network diagrams, normal arrows represent material transformation, rounded arrows represent catalysis, flat arrows represent inhibition, and \(\varnothing\) is a material sink. All food molecules \(F\) are constantly replenished and cannot be exhausted. \textbf{(a)} The single cell's physiology is composed of two interdependent molecules \(M_1\) and \(M_2\) which catalyze each other's synthesis from their respective food precursor. Both molecules decay into inconsequential waste products. \textbf{(b)} In the model of heterogeneous cells interacting, each autocatalytic molecule \(M_i\) represents the essential variable for one of two cells. These molecules are synthesized from the same food source and decay. It is assumed that the decay of each molecule releases a quickly diffusing and unstable byproduct that impacts the other molecule's autocatalysis (gray edges). \textbf{(c)} Eqns. 51 and 52 establish two opposing food gradients that the cell system defined in Fig. (a) can navigate once behavior is incorporated into the model.}
\end{figure}

These constraints form a viability region in the form of a triangle (light blue in Fig. 5(a)), which can then be broken into three viability boundary segments of codimension 1, which make three sharp junctures of codimension 2 where two constraints meet (Eqn. 12):
\begin{equation}
v_\alpha^1=\{\mathbf{x}\in\partial\mathcal{V}_\alpha:[M_1]=0.1\}
\end{equation}
\begin{equation}
v_\alpha^2=\{\mathbf{x}\in\partial\mathcal{V}_\alpha:[M_2]=0.1\}
\end{equation}
\begin{equation}
v_\alpha^3=\{\mathbf{x}\in\partial\mathcal{V}_\alpha:[M_1]+[M_2]=20\}
\end{equation}
This implies there are guard conditions specifying that the cell can die if its state strikes any of these three segments with outward velocity (Eqns. 6 and 13):
\begin{equation}
\mathcal{G}_\alpha^1=\{\mathbf{x}\in v_\alpha^1:[-1,0]\cdot[\dot{[M_1]},\dot{[M_2]}]^\top>0\}
\end{equation}
\begin{equation}
\mathcal{G}_\alpha^2=\{\mathbf{x}\in v_\alpha^2:[0,-1]\cdot[\dot{[M_1]},\dot{[M_2]}]^\top>0\}
\end{equation}
\begin{equation}
\mathcal{G}_\alpha^3=\{\mathbf{x}\in v_\alpha^3:[1,1]\cdot[\dot{[M_1]},\dot{[M_2]}]^\top>0\}
\end{equation}
This means that the model has six distinct possible ways to die, accounting for both the boundary segments and their intersections where multiple guard conditions could be satisfied. Accordingly, before pruning the agent-graph's structure has three solid edges for the possibility of a single constraint being violated, and three dashed edges for the possibility of two constraints being simultaneously violated (Fig. 5(e)).

\begin{figure*}
\includegraphics[width=6.0in]{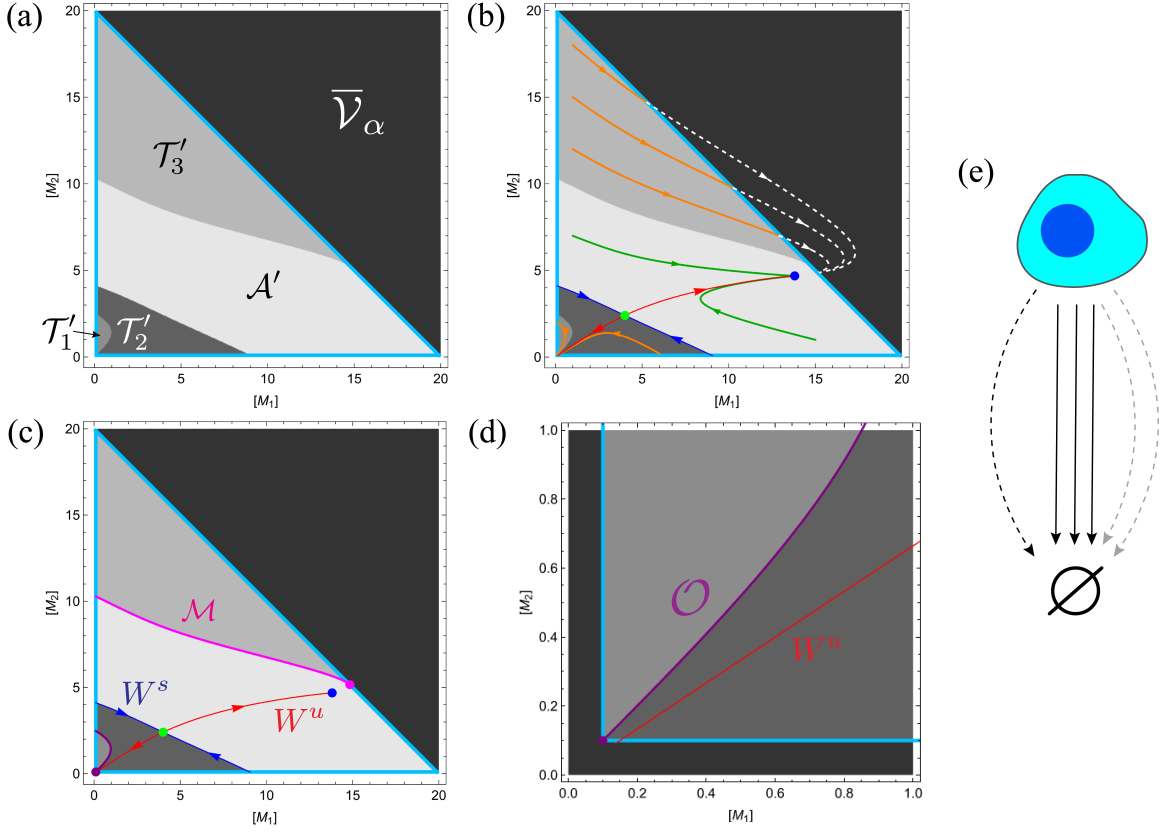}
\caption{\label{fig:epsart} \textbf{(a)} Sampling a million initial conditions in state space and simulating for a duration of 800 arbitrary time units, we color each initial condition according to its eventual fate, revealing five distinct regions. The first of these is the terminal region \(\overline{\mathcal{V}}_\alpha\) where the agent does not exist (black). This is trivially separated by the imposed viability boundary shown in light blue. Also present is one robust asymptotically viable region, and a robust transiently viable region corresponding to each of the three viability constraints. \textbf{(b)} A phase portrait analysis reveals a stable equilibrium point (dark blue) which asymptotically viable trajectories (dark green) converge to. Transiently viable initial conditions (orange) in \(\mathcal{T}'_3\) would converge to the attractor if the upper viability constraint did not exist (dashed white trajectories). The saddle node (light green) has two branches of \(W^u\) (red trajectories) which either reach the viable attractor or violate the \([M_2]=0.1\) constraint, resulting in \(W^s\) forming a boundary between \(\mathcal{A}'\) and \(\mathcal{T}'_2\). \textbf{(c)} Doing a complete viability space decomposition identifies the missing boundaries from Fig. (b). The mortality point (magenta) along the diagonal boundary represents where the change vector is tangent and then converges to the viable attractor. The backward time trajectory of the mortality point (magenta contour) is the mortality manifold that separates \(\mathcal{T}'_3\) from \(\mathcal{A}'\). The ordering point (purple) is where the dynamics result in violating both of the lower bound viability constraints simultaneously. Integrating it backward in time reveals the ordering manifold (purple contour) which separates \(\mathcal{T}'_1\) and \(\mathcal{T}'_2\). \textbf{(d)} Zooming into the lower left corner of the viability region makes it apparent that the branch of the unstable saddle manifold is part of \(\mathcal{T}'_2\), unlike the ordering manifold. \textbf{(e)} The geometry of the viability boundary allows us to translate the possible death transitions into an agent-graph. The two macro-states that the system can be in are the one where the cell is alive, and the one where it no longer exists. The cell's three viability constraints form three co-dimension 1 structures which can lead to its death, represented by solid arrows. The points where two constraints meet are co-dimension 2 structures that allow for the much rarer possibility of the cell dying by violating multiple constraints at once. The way the dynamics unfold within \(\mathcal{V}_\alpha\) mean only four of these six transitions are realized (dark edges).}
\end{figure*}

\subsubsection{Numerical exploration}

To approximate the possible existential outcomes in the single-cell physiology, we uniformly sample a million initial conditions in the range \([M_i]\in[0,20]\). We then numerically integrate each initial condition for up to 800 arbitrary time units (ATU), and color the initial condition according to whether it survives or violates one of the three viability boundary segments from Eqns. 43-48. This reveals five distinct regions. The complement of the viability region \(\overline{\mathcal{V}}_\alpha\), colored black, corresponds to where the total survival time is zero because the agent cannot exist beyond its viability region. The viability boundary \(\partial\mathcal{V}_\alpha\) trivially separates this portion from the other subsets where the agent persists for at least some time. Within the viability region, there are three robust transiently viable regions \(\mathcal{T}'_1\), \(\mathcal{T}'_2\), \(\mathcal{T}'_3\) which correspond to initial conditions that violate the \(v_\alpha^1\), \(v_\alpha^2\), and \(v_\alpha^3\)  viability boundary segments respectively. Finally there is one region where the system survives the whole time, which indicates a robust asymptotically viable region \(\mathcal{A}'\) where the cell achieves a homeostatic regime.

\subsubsection{Viability portrait}

A phase portrait analysis within the cell's viability region reveals two equilibrium points: a saddle node \(\mathrm{EP}_\mathcal{S}=(4.008,2.407)\) (light green point) and a stable equilibrium point \(\mathrm{EP}_\Omega=(13.811,4.686)\) (dark blue point), shown in Fig. 5(b). Sampling the flow, we can see that the asymptotically viable trajectories (green) converge to \(\mathrm{EP}_\Omega\), with the existence of \(\mathcal{A}'\) being dependent on its presence. Meanwhile, the orange trajectories show a sample of the dynamics in each of the three transiently viable regions, violating the three segments of the viability boundary \(v_\alpha^1\), \(v_\alpha^2\), and \(v_\alpha^3\) (Eqns. 43-45) by meeting their respective guard conditions \(\mathcal{G}_\alpha^1\), \(\mathcal{G}_\alpha^2\), and \(\mathcal{G}_\alpha^3\) (Eqns. 46-48). Looking in \(\mathcal{T}'_3\), we see that if the trajectories in this region had been able to continue past their collision with \(v_\alpha^3\) (dashed white trajectories), they also would have been part of  \(\mathrm{EP}_\Omega\)'s basin of attraction. 

To begin identifying the organizing manifolds of \(\mathcal{V}_\alpha\), we find the stable and unstable manifolds of \(\mathrm{EP}_\mathcal{S}\). Starting with the unstable manifold  \(W^u\) (red trajectories), one branch converges to  \(\mathrm{EP}_\Omega\) and one violates \(v_\alpha^2\), making them parts of \(\mathcal{A}'\) and \(\mathcal{T}'_2\) respectively. This means that the stable manifold \(W^s\) (blue trajectories) functions as the boundary between these two regions. This leaves us with two missing boundaries, which are needed to separate \(\mathcal{A}'\) from \(\mathcal{T}'_3\) and \(\mathcal{T}'_1\) from \(\mathcal{T}'_2\). Finding these missing boundaries requires us to carefully analyze how the geometry of \(\partial\mathcal{V}_\alpha\) interacts with the dynamics. 

The first step is to solve for the points where the change vector is tangent to the viability boundary, as described by Eqn. 16, since these points separate portions of \(\partial\mathcal{V}_\alpha\) that result in immediate death from those that instantaneously move back into the interior. This reveals three such points distributed across the viability boundary segments, two of which could be solved for analytically:
\begin{equation}
\Lambda= \Bigl\{ \left(\frac{1}{10}, \frac{4}{223^{1/3}}\right),\left(\frac{4}{47^{1/3}},\frac{1}{10}\right),(14.836,5.164)\Bigl\}
\end{equation}
Of these points, only the third remains asymptotically viable under the flow, making it a mortality point \(\mathbf{m}\) according to Eqn. 17 (magenta point in Fig. 5(c)). As detailed in Eqn. 18, the backward time trajectory of \(\mathbf{m}\) is a mortality manifold \(\mathcal{M}\) that separates the \(\mathcal{A}'\) and \(\mathcal{T}'_3\) regions (magenta contour), functioning as the final boundary needed to characterize them both as disjoint open sets. 

At this stage, we have fully decomposed \(\mathcal{V}_\alpha\), based on which initial conditions robustly converge to \(\mathrm{EP}_\Omega\) and remain asymptotically viable, but we have yet to explain the boundaries between transiently viable regions that lead to different death outcomes. Doing so involves solving for ordering points along the boundary that simultaneously satisfy the guard conditions of multiple viability segments, as in Eqn. 21. In our system this can be solved for analytically (purple point in Figs. 5(c) and (d)):
\begin{equation}
\mathcal{G}_\alpha^1\cap\mathcal{G}_\alpha^2=\{\mathbf{o}\}=  \Bigl\{ \left(\frac{1}{10},\frac{1}{10}\right)\Bigl\}
\end{equation}
The reverse-time trajectory of this ordering point (Eqn. 22) gives an ordering manifold \(\mathcal{O}\) shown as a purple contour that separates \(\mathcal{T}'_1\) and \(\mathcal{T}'_2\) (Fig. 5(c)). Zooming in closer around \(\mathbf{o}\) and sampling another 160,000 initial conditions for improved resolution, we can clearly see that \(\mathcal{O}\) separates the two regions while the \(W^u\) branch belongs to \(\mathcal{T}'_2\) (Fig. 5(d)). This final manifold finishes carving out the different regions of robust survival outcomes in \(\mathcal{V}_\alpha\). At this point, we can also prune edges in the agent-graph based on the fact that the guard conditions for \(v_\alpha^1\), \(v_\alpha^2\), and \(v_\alpha^3\) are all met at different regions of state space (solid black edges) and only one of the three possible points where constraint segments meet has both guard conditions active (one dashed black edge, two grayed out) (Fig. 5(e)). 

\subsection{Single-cell physiology and behavior}

\subsubsection{Model specification}

Our behaving cell model has the same physiology and viability constraints as the previous model, with the main difference being that we now allow the cell to move in a one-dimensional environment,  \(X\), which has opposing gradients of food concentrations, established by the following equations and illustrated in Fig. 4(c):   
\begin{equation}
[F_1] =  \frac{10}{0.2 + e^{-0.4 \cdot X - 0.22}}
\end{equation}
\begin{equation}
[F_2] = \frac{10}{0.2 + e^{0.4 \cdot X - 0.22}}
\end{equation}
The cell's movement in this environment is governed by a third differential equation that ensures the cell always moves toward the food that is a precursor for the lowest \([M_i]\). We do this by using a modification of an inhibitory Hill function that assesses the difference between \([M_1]\) and \([M_2]\), making sure that the cell moves to the right after \([F_1]\) when \([M_1]\) is lower (positive velocity) and  to the left after \([F_2]\) when \([M_2]\) is lower (negative velocity). The \(r_\mathrm{spatial}\) parameter then scales the maximum possible velocity, and all other parameters have the same interpretation as they did in the previously presented Hill functions. This behavior equation introduces a very slight asymmetry into the dynamics (see Appendix), but it is small enough that it can be mostly ignored for the sake of this paper:  
\begin{equation}
\frac{dX}{dt} = r_\mathrm{spatial} \left( \frac{K_2^{h_2}}{K_2^{h_2} + \left(\frac{[M_1] - [M_2]}{2([M_1] + [M_2])} + \frac{1}{2}\right)^{h_2}} -0.5 \right) \end{equation}
Since the behaving cell model has the same viability constraints as the model without behavior, the only difference is that \(\mathcal{V}_\alpha\) will be spread uniformly across the environmental dimension. This also means that the cell still has three boundary segments of codimension 1 that meet at codimension 2 intersections. The guard conditions are also the same, except they will have a zero entry for the environmental dimension. Before pruning, the agent-graph has the same structure as the model without behavior.

\subsubsection{Numerical exploration}

\begin{figure*}[p]
\includegraphics[width=6.4in]{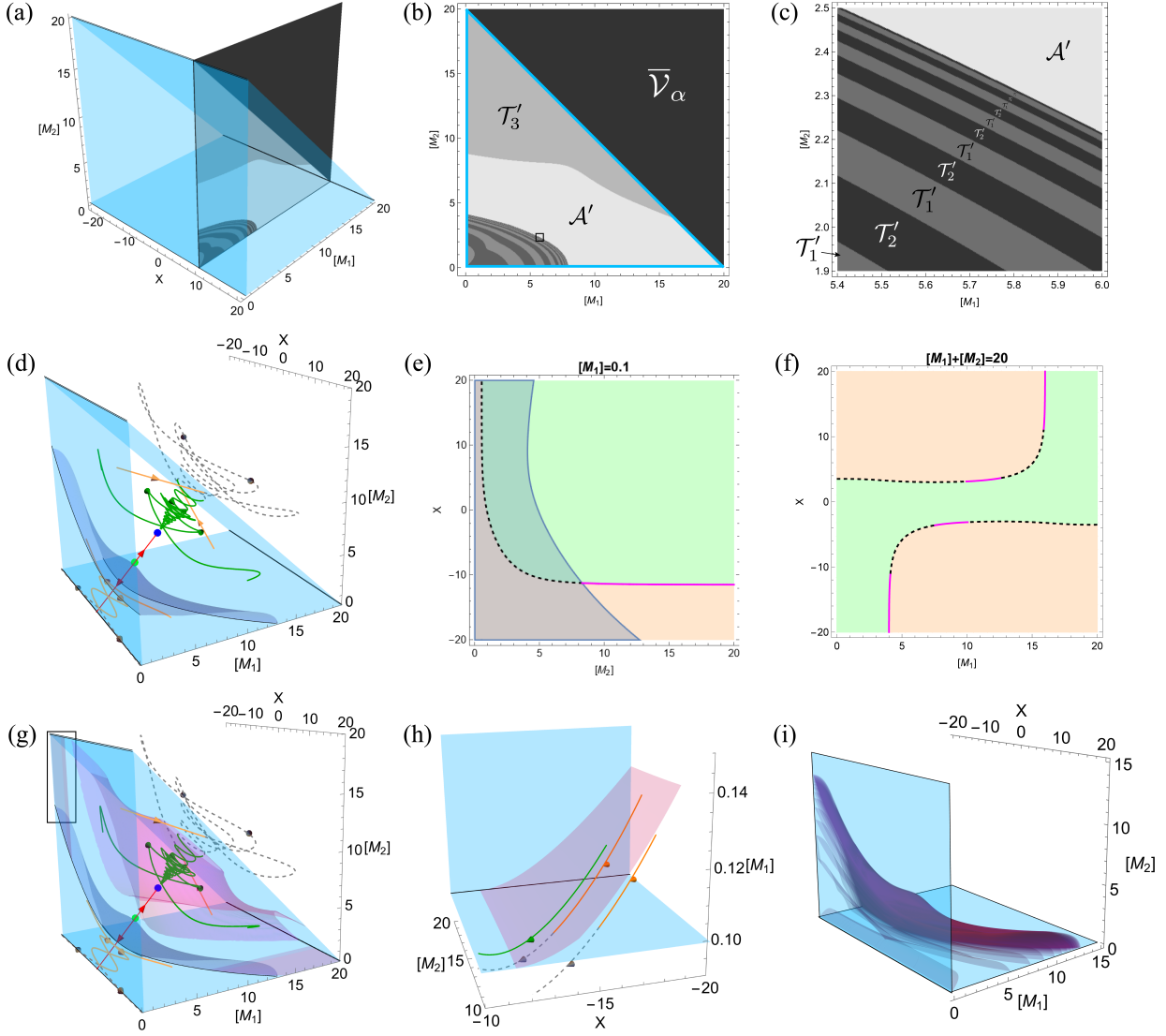}
\caption{\label{fig:epsart} \textbf{(a)} Including the environment dimension \(X\) spreads the triangular viability boundary out without altering its structure. We can then begin to understand the structure of survival outcomes in the space by taking a slice of a million initial conditions at \(X=10\) and shading them according to the survival outcomes. \textbf{(b)} Pulling out the slice from Fig. (a), we see five distinct outcomes. There is one robust asymptotically viable region \(\mathcal{A}'\) and a robust transiently viable region \(\mathcal{T}'_3\) where \(v_\alpha^3\) is violated. In the bottom left, there are alternating bands of regions where either \(v_\alpha^1\) or \(v_\alpha^2\) is violated. \textbf{(c)} Zooming in on the small black square in Fig. (b), where the plot resolution fails, another 160,000 initial conditions reveal that the alternating shrinking bands of  \(\mathcal{T}'_1\) and  \(\mathcal{T}'_2\) terminate. \textbf{(d)} The phase portrait analysis reveals a stable equilibrium point (dark blue) that asymptotically viable trajectories (dark green) converge to, and a saddle node (light green) whose stable manifold (dark blue) separates the transiently viable trajectories (orange) that correspond to the alternating bands of  \(\mathcal{T}'_1\) and  \(\mathcal{T}'_2\) from \(\mathcal{A}'\). We also see that the transiently viable trajectories originating in \(\mathcal{T}'_3\) would reach the attractor if they were not constrained by \(v_\alpha^3\) (upper right dashed gray trajectories). \textbf{(e)} Calculating the tangent vectors on \(v_\alpha^1\) and \(v_\alpha^2\) results in two nearly symmetric plots, with a tangency contour separating areas where vectors lead to immediate death (light orange) from areas where the state moves instantaneously into the interior (light green). The magenta portion of the tangency contour is asymptotically viable and therefore a set of mortality points, and the dashed black portion is transiently viable. The shaded blue region corresponds to the area on the transiently viable side of the stable saddle manifold. \textbf{(f)} The tangency calculations for \(v_\alpha^3\) with the same interpretation as Fig. (e). \textbf{(g)} Integrating the sets of mortality points backward in time reveals three mortality manifolds (magenta surfaces) that separate asymptotically and transiently viable initial conditions. The two smaller mortality manifolds appear almost flat and reveal transiently viable regions that are not easily detectable in the initial numerical sweep. \textbf{(h)} Zooming in on the mortality manifold in the black box in Fig. (g) shows that it is not flat and separates asymptotically and transiently viable trajectories. \textbf{(i)} The intersection of  \(v_\alpha^1\) and \(v_\alpha^2\) satisfies the guard conditions for both constraints, making it a set of ordering points. Integrating this set backwards in time gives us the ordering manifold (purple), separating \(\mathcal{T}'_1\) and  \(\mathcal{T}'_2\), and making it clear that the "bands" of similar fates we saw in Figs. (b) and (c) are connected.}
\end{figure*}

As with the cell physiology in a fixed environment, we approximate the possible existential outcomes for the behaving cell by sampling a million initial conditions spanning \([M_i]\in[0,20]\) simulated for 200 ATU, except now for slices of the environment \(X=[-20,-15,-10,-5,0,5,10,15,20]\). One representative slice is shown in Fig. 6(a) and then extracted for clarity in Fig. 6(b), with shading following the same scheme as in the previous model. The full set of slices can be seen in the Appendix. These scans reveal a robust transiently viable region \(\mathcal{T}'_3\) where the \(v_\alpha^3\) constraint is violated (Eqns. 45 and 48), and a single robust asymptotically viable region \(\mathcal{A}'\). In the lower left corner of Fig. 6B, we see what appears to be alternating shrinking bands of transiently viable initial conditions where either \(v_\alpha^1\) or \(v_\alpha^2\) is being violated (Eqns. 43, 44, 46, and 47). Fig. 6(c) zooms in on the small black square in Fig. 6(b) where the plot resolution begins to fail, and samples another 160,000 initial conditions, which are simulated for 200 ATU. This reveals that, while the alternating bands of \(\mathcal{T}'_1\) and \(\mathcal{T}'_2\) get progressively thinner as they approach \(\mathcal{A}'\), the pattern is not fractal and does eventually terminate. Notably, because we are looking at a two-dimensional slice through three-dimensional space, bands that violate the same viability constraint are not necessarily disconnected. Furthermore, which side of the environment we are on determines whether \(\mathcal{T}'_1\) or \(\mathcal{T}'_2\) borders \(\mathcal{A}'\). 

\subsubsection{Viability Portrait}

Similar to the protocell analysis without behavior, a phase portrait analysis within \(\mathcal{V}_\alpha\) reveals a stable equilibrium point \(\mathrm{EP}_\Omega=(6.434,6.434, 0)\) (dark blue) and a saddle node \(\mathrm{EP}_\mathcal{S}=(4.020,4.020, 0)\)  (light green) (Fig. 6(d)). The branches of the saddle node's one-dimensional unstable manifold \(W^u\) (red trajectories) either converge to \(\mathrm{EP}_\Omega\) as part of \(\mathcal{A'}\) or exactly terminate at the intersection of \(v_\alpha^1\) and \(v_\alpha^2\) at \((0.1,0.1,0)\) where the two corresponding guard conditions \(\mathcal{G}_\alpha^1\) and \(\mathcal{G}_\alpha^2\) are simultaneously satisfied. Accordingly, the saddle node's two-dimensional stable manifold (dark blue) separates the bands of \(\mathcal{T}'_1\) and \(\mathcal{T}'_2\) from \(\mathcal{A}'\) (Fig. 6(c)). 

Solving for mortality manifolds once again requires us to start by solving for tangent change vectors along the viability boundary segments, except that now we need to solve for one-dimensional contours rather than isolated points. For \(v_\alpha^1\) and \(v_\alpha^2\), this means looking for where the rate of change of the constrained chemical species is zero (Eqn. 16):
\begin{equation}
\Lambda_1= \{ \mathbf{x}\in v_\alpha^1:\dot{[M_1]}=0\}
\end{equation}
\begin{equation}
\Lambda_2= \{ \mathbf{x}\in v_\alpha^2:\dot{[M_2]}=0\}
\end{equation}
Since the model is symmetric with respect to both the cell circuit and environmental gradients, the contours on each of these constraints are reflections of one another, so we only show one in Fig. 6(e). As expected, we can see that the tangency contour separates areas on the viability boundary segment where the change vector is terminal (light orange) or instantaneously recovers (light green). 

Remaining tangent to the \(v_\alpha^3\) boundary segment requires finding where the sum of the chemical concentrations does not change, which gives two disconnected tangency contours that once again separate the areas where \(\mathcal{G}_\alpha^3\) is satisfied from where the change vector points back into the interior:
\begin{equation}
\Lambda_3= \{ \mathbf{x}\in v_\alpha^3:\dot{[M_1]}+\dot{[M_2]}=0\}
\end{equation}
Within these tangency contours, the only points that participate in the global organization of \(\mathcal{V}_\alpha\) are the mortality points that remain asymptotically viable (Eqn. 17). This requires converging to the system's only viable attractor,  \(\mathrm{EP}_\Omega\)  (magenta in Figs. 6(e) and (f)):
\begin{equation}
\mathbf{M}_i= \{ \mathbf{x}\in \Lambda_i:\phi_{t\rightarrow\infty}(\mathbf{x})=\mathrm{EP}_\Omega\}
\end{equation}
According to Eqn. 19, the reverse-time flow of each family of mortality points is a mortality manifold \(\mathcal{M}_i\) (magenta boundaries in Fig. 6(g)):
\begin{equation}
\mathcal{M}_i=\bigcup_{\mathbf{x}\in\mathbf{M}_i} \{ \phi_{t}(\mathbf{x}):t \leq 0\}
\end{equation}
Of these, \(\mathcal{M}_3\) functions as the missing boundary between \(\mathcal{A}'\) and \(\mathcal{T}'_3\), but \(\mathcal{M}_1\) and \(\mathcal{M}_2\) actually carve out two additional transiently viable regions that border \(\mathcal{A}'\) but are small enough that they are not easily detectable by just doing a numerical sweep. Zooming in on \(\mathcal{M}_2\) in the black square from Fig. 6(g) reveals that these manifolds do not actually press flat against the viability boundary segments (Fig. 6(h)).

At this stage we have identified all of the boundaries that carve up state space based on which initial conditions will perish and which will remain asymptotically viable, but we have not explained what separates the \(\vn{\mathcal{T}'_1}\) and \(\vn{\mathcal{T}'_2}\) bands in Fig 6(c). Looking at where \(v_\alpha^1\) and \(v_\alpha^2\) meet, we see both of their guard conditions are simultaneously met, making this a set of ordering points \(\mathbf{O}\):
\begin{equation}
\mathbf{O}=\mathcal{G}_\alpha^1 \cap \mathcal{G}_\alpha^2 = v_\alpha^1 \cap v_\alpha^2
\end{equation}
The missing ordering manifold \(\mathcal{O}\) is the backward time flow of this set, as specified in Eqn. 23 (purple boundary in Fig. 6(i)). The structure of \(\mathcal{O}\) reveals several important details about the robust transiently viable regions \(\mathcal{T}'_1\) and \(\mathcal{T}'_2\). For starters, the bands of similar death outcomes we saw in Figs. 6(b) and (c) are not actually disconnected. Instead, \(\mathcal{O}\) essentially twists around its center, the branch of the unstable saddle manifold \(W^u\) that is within its set. These twists effectively funnel the \(\mathcal{T}'_1\) and \(\mathcal{T}'_2\) families of trajectories from side to side until they eventually satisfy the respective \(\mathcal{G}_\alpha^1\) or \(\mathcal{G}_\alpha^2\) guard condition. 
\begin{figure}[t]
\includegraphics[width=\columnwidth]{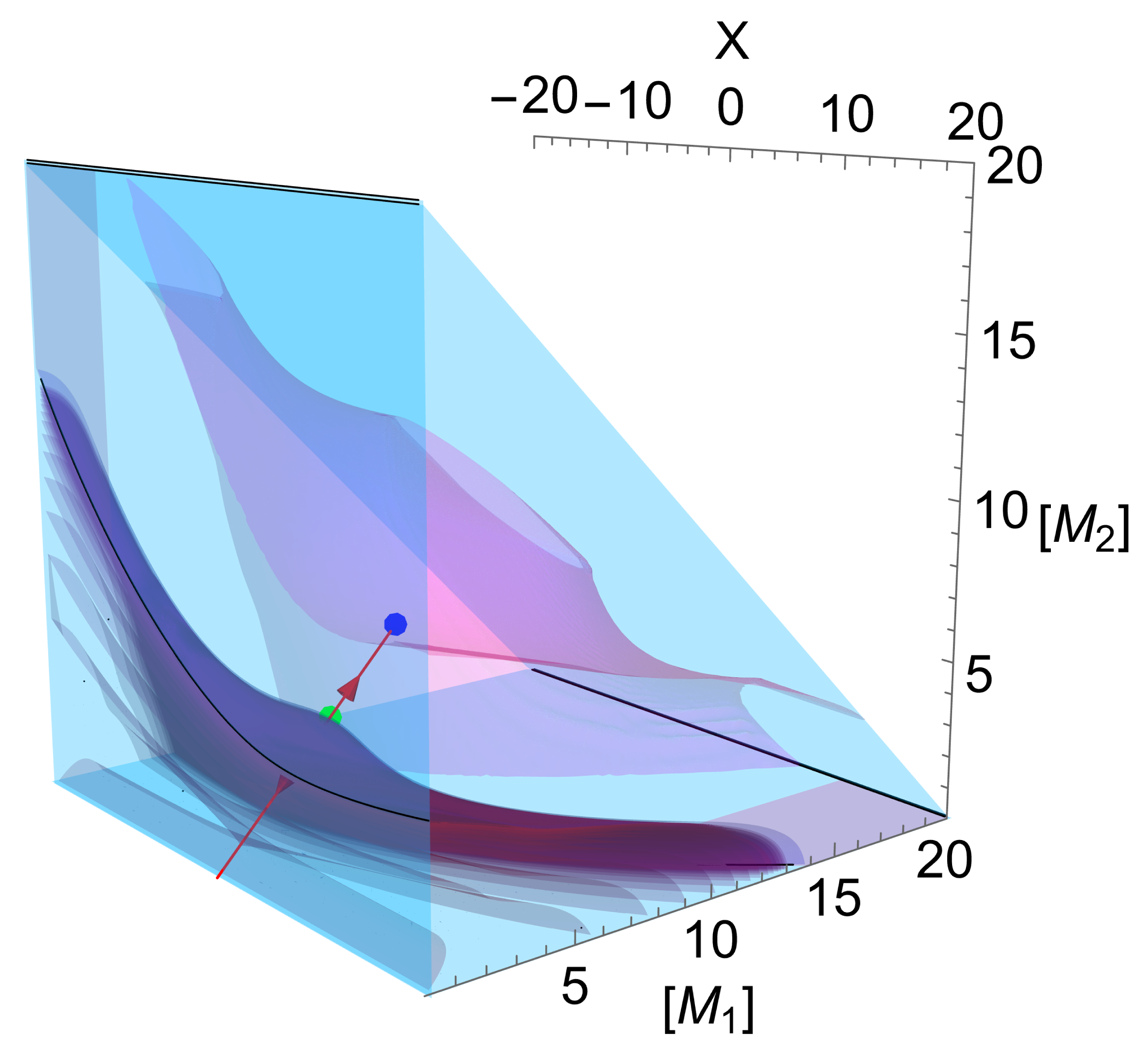}
\caption{\label{fig:epsart} Combining all of the elements from Fig. 6 gives us the behaving cell's complete viability portrait.}
\end{figure}
Identifying these limit sets and global manifolds culminates in the viability portrait for our behaving cell (Fig. 7). Since there are points where each of the viability boundary segments are violated and one intersection where both \(v_\alpha^1\) and \(v_\alpha^2\) are violated, the agent-graph ends up having the same structure as in Fig. 5(e).

\subsection{Two heterogeneous cells interacting}

\subsubsection{Model specification}
Like the single-cell physiology model, our multi-agent model will be comprised of two cells, \(\alpha=\{\alpha_1,\alpha_2\}\), where \([M_1]\) and \([M_2]\) belong to each of these agents respectively (Fig. 4(b)). We assume that these molecules are synthesized from the same precursor food molecule, but beyond this the synthesis is the same as in Eqn. 38:  
\begin{equation}
S([M_i]) = r_{M_i}\left( \frac{[M_{i}]^{h_1}}{K_1^{h_1}+[M_i]^{h_1}} \right) [F]
\end{equation}
The rate of decay of each molecule is determined by the same linear function from Eqn. 39. We also introduce the idea that as \([M_i]\) decays, it releases an unstable and quickly diffusing byproduct that influences the other cell's rate of synthesis, coupling the two cells' dynamics. Each equation \(I_{i,j}\) describes the impact that \(\alpha_i\) has on \(\alpha_j\), where each takes the form of a Hill function when the other cell is present. Eqn. 61 specifies that \(\alpha_2\) has an excitatory impact on \(\alpha_1\), and Eqn. 62 that \(\alpha_1\) has the opposite impact on  \(\alpha_2\). The parameters \(d_i\) are dimensionless scalars:
\begin{equation}
I_{2,1}([M_2]) = \begin{cases} \frac{(d_2 \cdot D([M_{2}]))^{h_2}}{K_2^{h_2}+(d_2 \cdot D([M_{2}]))^{h_2}} & \text{if  }\alpha_2\\ 1 & \text{otherwise} \end{cases}
\end{equation}
\begin{equation}
I_{1,2}([M_1]) = \begin{cases} \frac{K_2^{h_2}}{K_2^{h_2}+(d_1 \cdot D([M_{1}]))^{h_2}} & \text{if  }\alpha_1\\ 1 & \text{otherwise} \end{cases}
\end{equation}
These functions taken together give us the dynamics of the cells, both when they are coupled and when they are found in isolation of one another: 
\begin{equation}
\frac{d[M_1]}{dt} = S([M_1]) \cdot I_{2,1}([M_2]) - D([M_1])
\end{equation}
\begin{equation}
\frac{d[M_2]}{dt} = S([M_2]) \cdot I_{1,2}([M_1]) - D([M_2])
\end{equation}
Similar to the single-cell model, each cell's viability region has viability constraints that determine when molecular concentrations become too low and when they are high enough to cause an osmotic crisis. In this case that means each \(\alpha_i\) having an upper and lower bound on \([M_i]\): 
\begin{equation}
\mathcal{V}_{\alpha_i}\triangleq  \{ \mathbf{x}\in \mathcal{U}:[M_i]\geq0.1 \wedge [M_i]\leq10 \}
\end{equation}
This means each cell's viability region is contained by two disconnected boundary segments: 
\begin{equation}
v_{\alpha_i}^1=\{\mathbf{x}\in\partial\mathcal{V}_{\alpha_i}:[M_i]=0.1\}
\end{equation}
\begin{equation}
v_{\alpha_i}^2=\{\mathbf{x}\in\partial\mathcal{V}_{\alpha_i}:[M_i]=10\}
\end{equation}
The two cells then have a domain of interaction where all their constraints are mutually satisfied: 
\begin{equation}
\mathcal{X}\triangleq \bigcap \{ \mathcal{V}_{\alpha_1},\mathcal{V}_{\alpha_2} \}
\end{equation}
While each cell's viability boundary segments are disconnected and correspond to codimension 1 structures where they could die, there are four codimension 2 points where the viability boundary segments of the two cells intersect. This means that the agent-graph will have two outgoing solid edges for each possible death transition where a single cell could die, and four dashed edges for the cases where two cells might die simultaneously (Fig. 3(b)).

Defining the guard conditions in this model is more complicated than in the single-cell models because the possibility of collapsing into a lower-dimensional space means we cannot keep comparing velocity against fixed-dimensional orthogonal vectors at the viability boundary segments. To get around this, we can define an oriented distance function for an agent \(\alpha_i\)'s state relative to its viability segment \(v_{\alpha_i}^j\) such that: 
\begin{equation}
d_{\alpha_i}^{v_{\alpha_i}^j}(\mathbf{x}) = \begin{cases} >0 & \text{if  }\mathbf{x}\text{ is on the viable side of }{v_{\alpha_i}^j}\\ 0 & \text{if }\mathbf{x}\in v_{\alpha_i}^j \\ <0 & \text{if }\mathbf{x}\text{ is on the nonviable side of }{v_{\alpha_i}^j} \end{cases}
\end{equation}
When we are at the viability boundary segment, the time derivative of the oriented distance function at that point will be positive when the change vector pulls the state away, zero when the change vector is tangent, and negative when the change vector is striking the viability boundary with outward velocity: 
\begin{equation}
\frac{d}{dt}d_{\alpha_i}^{v_{\alpha_i}^j}=\nabla d_{\alpha_i}^{v_{\alpha_i}^j}\cdot \mathbf{F}(\mathbf{x})
\end{equation}
Accordingly, we can define the guard conditions for an agent dying via violating particular constraints as:
\begin{equation}
\mathcal{G}_{\alpha_i}^j=\Bigl\{ \mathbf{x}\in v_{\alpha_i}^j:\frac{d}{dt}d_{\alpha_i}^{v_{\alpha_i}^j}<0\Bigl\}
\end{equation}
Finally, we need to declare the collapse functions that define how the model resets when guard conditions are satisfied, as specified in Eqns. 29 and 30. If one cell \(\alpha_i\) satisfies its guard conditions for death in \(\mathcal{X}\), this just involves removing its state dimension such that the surviving cell \(\alpha_j\)'s dynamics unfold in its uncoupled viability region:
\begin{equation}
\Downarrow_{\alpha_i}|_{\mathcal{G}_{\alpha_i}}:\mathcal{X}\rightarrow\mathcal{V}_{\alpha_j}
\end{equation}
Since there are only two agents in this model, if an isolated cell dies within its viability region, or if both cells meet their guard conditions for death simultaneously within \(\mathcal{X}\), the collapse functions will transition us to the world state where the agent list is empty, effectively terminating the model. This gives the hybrid automaton illustrated in Fig. 3(c). 

\subsubsection{Numerical exploration}

We sample a million initial conditions within the domain of interaction of the two cells \(\mathcal{X}\) and then simulate their trajectories for 800 ATU, revealing six disconnected regions of qualitatively distinct survival outcomes (Fig. 8(a)). Regions are given labels according to their sequence of asymptotically and transiently viable outcomes. When referring to a transiently viable outcome, the superscript marks the domain that the death event occurs in, and the subscript refers to the agent that has died. Asymptotically viable outcomes only mark the domain in which they occur since any homeostatic regime achieved involves all participating agents and cannot be attributed to a single one of them. If a single agent achieves asymptotic viability, this is implicitly accounted for by representing the outcome occurring in a viability region \(\mathcal{V}_{\alpha_i}\) as opposed to a domain of interaction. 

In accord with this notation, the \(\vn{\mathcal{T}^\mathcal{X}_{\alpha_1}\mathcal{A}^{\mathcal{V}_{\alpha_2}}}\) region in the top left of Fig. 8(a) corresponds to trajectories with the survival sequence where the cell \(\alpha_1\) dies from violating the \(v_{\alpha_1}^1\) viability boundary segment in \(\mathcal{X}\), followed by \(\alpha_2\) surviving and achieving an asymptotically viable state in \(\mathcal{V}_{\alpha_2}\). To the right of this is an asymptotically viable region \(\vn{\mathcal{A}^\mathcal{X}}\) where both cells cooperatively achieve a homeostatic regime and remain asymptotically viable within \(\mathcal{X}\). The \(\vn{\mathcal{T}^\mathcal{X}_{\alpha_1}\mathcal{A}^{\mathcal{V}_{\alpha_2}}}\) region to the right of \(\vn{\mathcal{A}^\mathcal{X}}\) has the same sequence of outcomes as the one to the left, except here \(\alpha_1\) dies after violating its \(v_{\alpha_1}^2\) boundary segment. Moving to the lower left, \(\vn{\mathcal{T}^\mathcal{X}_{\alpha_1}\mathcal{T}_{\alpha_2}^{\mathcal{V}_{\alpha_2}}}\) is characterized by \(\alpha_1\) violating \(v_{\alpha_1}^1\) in \(\mathcal{X}\) followed by \(\alpha_2\) violating \(v_{\alpha_2}^1\) in \(\mathcal{V}_{\alpha_2}\). The final two regions of survival outcomes are both labeled \(\vn{\mathcal{T}^\mathcal{X}_{\alpha_2}\mathcal{T}^{\mathcal{V}_{\alpha_1}}_{\alpha_1}}\) and involve \(\alpha_2\) violating  \(v_{\alpha_2}^1\), but are distinguished by whether \(\alpha_1\) dies at its  \(v_{\alpha_1}^1\) or  \(v_{\alpha_1}^2\) boundary segment in \(\mathcal{V}_{\alpha_1}\). 

\subsubsection{Viability portrait}

\begin{figure*}[p]
\includegraphics[width=\textwidth]{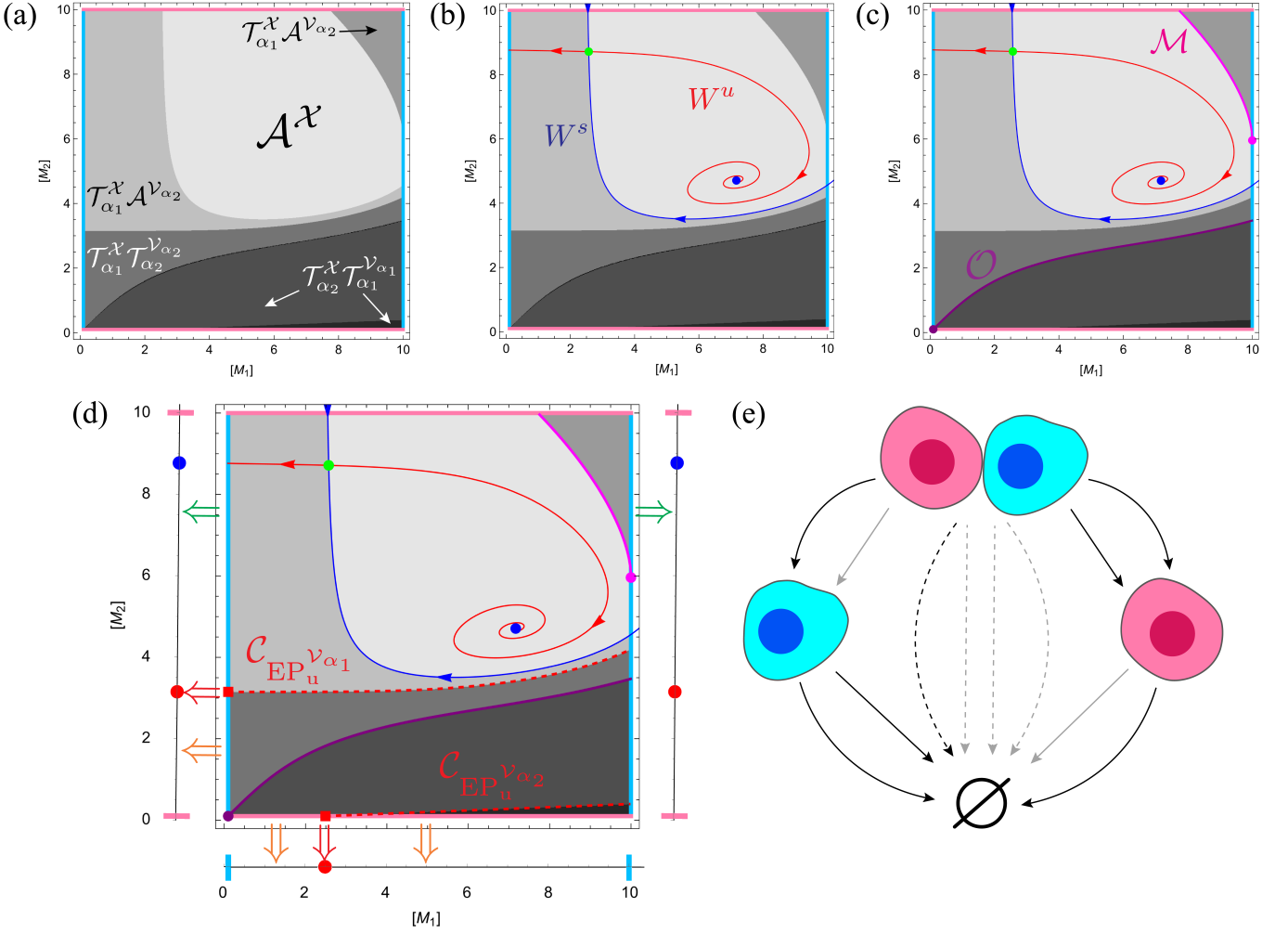}
\caption{\label{fig:epsart} \textbf{(a)} Running a million initial conditions for 800 ATU within the domain of interaction \(\mathcal{X}\) reveals six different regions of robust sequences of existential outcomes. \(\smash{\mathcal{T}^\mathcal{X}_{\alpha_1}\mathcal{A}^{\mathcal{V}_{\alpha_2}}}\) results in \(\alpha_1\) dying in the domain of interaction followed by an asymptotically viable regime in \(\alpha_2\)'s viability region. \(\mathcal{A}^\mathcal{X}\) initial conditions lead to all agents in the domain of interaction being asymptotically viable by cooperatively achieving a homeostatic regime. The second \(\vn{\mathcal{T}^\mathcal{X}_{\alpha_1}\mathcal{A}^{\mathcal{V}_{\alpha_2}}}\) results in the same sequence of existential events as the first, except now \(\alpha_1\) dies from violating its \(v^2_{\alpha_1}\) constraint. \(\vn{\mathcal{T}^\mathcal{X}_{\alpha_1}\mathcal{T}^{\mathcal{V}_{\alpha_2}}_{\alpha_2}}\) leads to \(\alpha_1\) dying in the domain of interaction followed by \(\alpha_2\) dying while alone in its viability region. Finally, there are two different \(\vn{\mathcal{T}^\mathcal{X}_{\alpha_2}\mathcal{T}^{\mathcal{V}_{\alpha_1}}_{\alpha_1}}\) regions bordering each other with the only difference being whether \(\alpha_1\) dies from violating \(v_{\alpha_1}^1\) or \(v_{\alpha_1}^2\). \textbf{(b)} The phase portrait within \(\mathcal{X}\) reveals a stable equilibrium point (dark blue) and a saddle node (light green). The branches of the saddle node's unstable manifold \(W^u\) (red trajectories) are members of the \(\vn{\mathcal{T}^\mathcal{X}_{\alpha_1}\mathcal{A}^{\mathcal{V}_{\alpha_2}}}\) and \(\mathcal{A}^\mathcal{X}\) sets, resulting in the stable saddle manifold \(W^s\) (dark blue trajectories) forming a boundary between these regions. \textbf{(c)} Using the same tools as in single-agent viability space decomposition, we can find the missing manifolds that organize regions based on similar survival outcomes within \(\mathcal{X}\). This gives us a mortality manifold \(\mathcal{M}\) (magenta contour) which separates \(\mathcal{A}^\mathcal{X}\) from the \(\vn{\mathcal{T}^\mathcal{X}_{\alpha_1}\mathcal{A}^{\mathcal{V}_{\alpha_2}}}\) region to the right, and an ordering manifold \(\mathcal{O}\) (purple contour) that separates \(\vn{\mathcal{T}^\mathcal{X}_{\alpha_1}\mathcal{T}^{\mathcal{V}_{\alpha_2}}_{\alpha_2}}\) from one of the \(\vn{\mathcal{T}^\mathcal{X}_{\alpha_2}\mathcal{T}^{\mathcal{V}_{\alpha_1}}_{\alpha_1}}\) regions. \textbf{(d)} To finish the viability portrait for the two-cell system, we decompose the one-dimensional viability regions where the collapse functions bring the surviving cells. This reveals that \(\alpha_1\) has no attractor in \(\mathcal{V}_{\alpha_1}\) and an unstable equilibrium point (red point) that separates initial conditions in the space that will violate \(v^1_{\alpha_1}\) or \(v^2_{\alpha_1}\). \(\mathcal{V}_{\alpha_2}\) on the other hand has a region that converges to a stable equilibrium point (dark blue) and a region that violates \(v^1_{\alpha_2}\) separated by an unstable equilibrium point (red point). Once we have identified these lower-dimensional organizing structures, we look for which terminal points in \(\partial\mathcal{X}\) collapse onto the unstable equilibrium points (red squares). Integrating these points backward in time extends the unstable equilibria into \(\mathcal{X}\) as collapse manifolds (dashed red contours) which give us the two final boundaries needed for the viability portrait. \textbf{(e)} Based on the geometry of the viability boundaries and their interaction with the dynamics, the directed agent-graph has two solid edges for each node including \(\alpha_1\) (light blue cell), one solid edge for each node containing \(\alpha_2\) (light red cell), and one dashed edge for the single intersection of constraints where both die simultaneously.}
\end{figure*}

Decomposing our two-cell state space demands that we understand how our agents' dynamics unfold within the intersection of their viability regions \(\mathcal{X}\). A phase portrait analysis of the unified state space reveals a stable equilibrium point  \(\mathrm{EP}_\Omega^\mathcal{X}=(7.167,4.707)\) (dark blue point) and a saddle node \(\mathrm{EP}_\mathcal{S}^\mathcal{X}=(2.571,8.716)\) (light green) (Fig. 8B). Although homeostasis is usually thought of as being a property of individual organisms, the attractor  \(\mathrm{EP}_\Omega^\mathcal{X}\) that permits this is realized through the mutual behavior of the two cells, explaining how trajectories in \(\mathcal{A}^\mathcal{X}\) remain asymptotically viable. The unstable manifold \(W^u\) branches (red trajectories) of \(\mathrm{EP}_\mathcal{S}^\mathcal{X}\) fall within the \(\mathcal{T}^\mathcal{X}_{\alpha_1}\mathcal{A}^{\mathcal{V}_{\alpha_2}}\) region in the upper left and adjacent \(\mathcal{A}^\mathcal{X}\) region, such that the stable saddle manifold \(W^s\) (blue trajectories) forms the boundary between these two sets. This leaves four missing boundaries that still need to be identified.

As in the single-agent models we have already decomposed, ordering and mortality manifolds allow us to decompose \(\mathcal{X}\) according to how trajectories remain in or leave this bounded region. Solving for tangencies along each agent's viability boundary segment gives an analytical solution for a point along \(\alpha_1\)'s \(v^2_{\alpha_1}\) constraint (Eqn. 70).  Numerically integrating forward in time confirms that the trajectory is part of \(\mathcal{A}^\mathcal{X}\), making it a mortality point \(\mathbf{m}\) (magenta dot in Fig. 8(c)) (Eqn. 17):
\begin{equation}
\mathbf{m}= \left(10, \frac{4\cdot133^{1/4}}{3^{3/4}}\right) 
\end{equation}
The backward time trajectory of \(\mathbf{m}\) then gives us a mortality manifold \(\mathcal{M}\) (magenta contour in Fig. 8(c)) which separates  \(\mathcal{A}^\mathcal{X}\) from the \(\mathcal{T}^\mathcal{X}_{\alpha_1}\mathcal{A}^{\mathcal{V}_{\alpha_2}}\) in the upper right of \(\mathcal{X}\) (Eqn. 18). 

Moving on to the intersection of viability boundary segments, we can analytically identify that there is a single point where \(\alpha_1\) and \(\alpha_2\)'s guard conditions are both satisfied, giving us an ordering point \(\mathbf{o}\) (Eqn. 21): 
\begin{equation}
\mathcal{G}_{\alpha_1}^1 \cap \mathcal{G}_{\alpha_2}^1=\{\mathbf{o}\}= \Bigl\{ \left(\frac{1}{10},\frac{1}{10}\right)\Bigl\}
\end{equation}
The backward-time trajectory of \(\vn{\mathbf{o}}\) gives us the ordering manifold \(\mathcal{O}\) (purple contour in Fig. 8(c)) which separates \(\vn{\mathcal{T}^\mathcal{X}_{\alpha_1}\mathcal{T}^{\mathcal{V}_{\alpha_2}}_{\alpha_2}}\) from the bordering \(\vn{\mathcal{T}^\mathcal{X}_{\alpha_2}\mathcal{T}^{\mathcal{V}_{\alpha_1}}_{\alpha_1}}\) region, effectively separating the two transiently viable regions based on which agent will die within \(\mathcal{X}\) (Eqn. 22). 

At this stage, we have accounted for all of the boundaries that are needed to differentiate between the robust existential outcomes that occur within \(\mathcal{X}\), but we are still missing two boundaries based on the fates that befall the agents after these first existential events. In order to complete the decomposition, we need to account for the structure of their uncoupled viability regions, and then extend their organizing boundaries back into \(\mathcal{X}\) (Fig. 8(d)). Since the uncoupled cells are one-dimensional systems, a phase portrait analysis alone is enough to understand the global structure of their survival outcomes. In \(\vn{\mathcal{V}_{\alpha_1}}\), we find a single unstable equilibrium point \(\vn{\mathrm{EP}_\mathrm{u}^{\mathcal{V}_{\alpha_1}}=2.492}\) (red point) and no other limit sets, meaning that asymptotic viability is not possible and which side of \(\vn{\mathrm{EP}_\mathrm{u}^{\mathcal{V}_{\alpha_1}}}\) \(\alpha_1\) falls on will determine whether it violates \(v_{\alpha_1}^1\) or  \(v_{\alpha_1}^2\). \(\mathcal{V}_{\alpha_2}\) also has an unstable equilibrium point  \(\vn{\mathrm{EP}_\mathrm{u}^{\mathcal{V}_{\alpha_2}}=3.15}\) (red point), except now one side of it converges to a viable attractor  \(\vn{\mathrm{EP}_\Omega^{\mathcal{V}_{\alpha_2}}=8.767}\) (blue point). 

Mapping each of the unstable equilibria through the inverse of the missing agent's collapse function allows us to find the terminal point that falls exactly onto them (red squares). These terminal states can then be integrated backward in time, extending the unstable equilibria into the higher dimensional domain of interaction \(\vn{\mathcal{X}}\) as the collapse manifolds \(\vn{\mathcal{C}_{\mathrm{EP}_\mathrm{u}^{\mathcal{V}_{\alpha_1}}}}\) and \(\vn{\mathcal{C}_{\mathrm{EP}_\mathrm{u}^{\mathcal{V}_{\alpha_2}}}}\) (red dashed lines in Fig. 8(d)). Accordingly, \(\vn{\mathcal{C}_{\mathrm{EP}_\mathrm{u}^{\mathcal{V}_{\alpha_2}}}}\) ends up separating \(\vn{\mathcal{T}^\mathcal{X}_{\alpha_1}\mathcal{A}^{\mathcal{V}_{\alpha_2}}}\) and \(\vn{\mathcal{T}^\mathcal{X}_{\alpha_1}\mathcal{T}^{\mathcal{V}_{\alpha_2}}_{\alpha_2}}\) based on whether \(\Downarrow_{\alpha_1}\) leaves \(\alpha_2\) to converge to \(\vn{\mathrm{EP}_\Omega^{\mathcal{V}_{\alpha_2}}}\) (green double arrow to the left of \(\mathcal{X}\)) or violate \(v_{\alpha_2}^1\) (orange double arrow to the left of \(\mathcal{X}\)). For the \(\mathcal{T}^\mathcal{X}_{\alpha_1}\mathcal{A}^{\mathcal{V}_{\alpha_2}}\) region in the upper right, no terminal points map to the transiently viable region or the unstable equilibrium, which is why there is  no collapse manifold there. \(\vn{\mathcal{C}_{\mathrm{EP}_\mathrm{u}^{\mathcal{V}_{\alpha_1}}}}\) similarly separates the \(\vn{\mathcal{T}^\mathcal{X}_{\alpha_2}\mathcal{T}^{\mathcal{V}_{\alpha_1}}_{\alpha_1}}\) regions at the bottom of \(\mathcal{X}\) based on which side of  \(\vn{\mathrm{EP}_\mathrm{u}^{\mathcal{V}_{\alpha_1}}}\) the collapse function leaves \(\alpha_1\)'s state on.

Ultimately, these boundaries spanning combinations of viability regions give us a complete viability portrait for the two cell system. Based on the guard conditions activated for each configuration of agents, the overall death transitions can be summarized by the agent-graph depicted in Fig. 8(e), where the dark edges correspond to those transitions that are enabled by the dynamics.

\section{Discussion}

In this paper, we introduced viability space decomposition as an extension of dynamical systems theory, ultimately allowing us to identify global manifolds that separate distinct survival outcomes in ordinary differential equation models of agents with viability constraints. We developed this approach in a way that identified global organizing principles, described the corresponding manifolds for state spaces of arbitrary dimensionality, and extended this geometric approach to systems of multiple agents. This pushed us into a hybrid dynamical system formalism, where death is an event that punctuates continuous physiological and behavioral dynamics, and reforms the state space to match the new set of agents. We also demonstrated the theory in action by completely mapping the survival outcomes in three models of cells, exploring physiological subsystems, coupled physiology and behavior, and multi-agent interaction. Beyond predicting the existence of regions with distinct survival outcomes, perhaps the greatest advantage of this theory is that it explains how such regions emerge through the geometric interaction of viability constraints and flows, ultimately introducing new organizing structures that exist on the same level as limit sets and their corresponding invariant manifolds. 

While viability space decomposition aims to be a complete theory for the state space analysis of generic autonomous dynamical systems with viability constraints, certain slight modifications may be necessary to deal with particular models. Nonetheless, this can often be done without impacting the core results of the theory as presented. For example, our state spaces were instances of \(\mathbb{R}^n\), but there is no reason why we could not utilize more complex manifolds, as in standard dynamical systems theory, and define the viability regions relative to these spaces. Relatedly, while all of the systems in this paper utilized physiological components as essential variables, the same decomposition can be done while making an agent's viability constrained by something external, such as location in an oxygen gradient \cite{lowengrubNonlinearModellingCancer2010}. 

It is also possible to represent other types of existential transformations within the formalism. If we wanted to modify death to leave behind debris in the environment, such as how cells can leave behind previously intracellular components in a regulated or unregulated manner \cite{yuanGuideCellDeath2024}, the collapse function would still modify the list of agents, but also influence the state of the environment, possibly adding new dimensions. Transitions besides death are also possible. For example, cells undergoing mitosis would still involve the end of the individual in a sense, except now instead of a collapse in the agent-list there would be a ``lift function'' \((\Uparrow)\) that raises the system into a higher dimensional space. Existential transformations that do not end an agent but fundamentally alter it in some way, such as altering its governing equations, can also be described as discrete transitions in the hybrid automaton (technically such a transition was incorporated in our two-cell example due to the structure of Eqns. 61 and 62). As in the case of the collapse function, the inverse mapping of the reset functions could still be used to extend any organizing manifolds that originate in these new domains, assuming they have certain features.

In terms of applications, there are numerous domains where viability space decomposition could provide novel insights. For starters, the framework has been presented as a way to make progress on fate outcomes in computational cell biology \cite{mcshaffreyMattersLifeDeath2026}, building on the already present intuition in systems biology of cellular states unfolding in an abstract geometric space \cite{ferrellBistabilityBifurcationsWaddingtons2012, rajuGeometricalPerspectiveDevelopment2023}. This perspective also has potential synergy with recent single-cell longitudinal experiments which connect initial states to particular fate outcomes \cite{goyalDiverseClonalFates2023,weilerCellRank2Unified2024}, sometimes even constructing dynamical flows similar to those considered in this paper \cite{qiuMappingTranscriptomicVector2022}. While it has recently been proposed that dynamical systems theory may function as an organizing set of principles for these methods \cite{pillaiUnravelingNongeneticHeterogeneity2023, islamDynamicalSystemsTheory2025}, in this paper we have demonstrated that dynamical systems theory alone is not sufficient for understanding fate outcomes when viability is in question. As a result, it may be possible to utilize the global principles offered by viability space decomposition to derive new insights in these studies. Viability space decomposition can also play this role for models of multicellular organisms that assess dynamics relative to physiological constraints, including biomedical settings where we must consider issues of disease onset and severity \cite{voitSystemstheoreticalFrameworkHealth2009,davisDynamicalSystemsApproaches2019,alonSystemsMedicinePhysiological2023}. More generally though, the theory only depends on the flow of some dynamical system and the existence of a boundary that can be crossed. To this end, recent work has repurposed the ideas presented here to better predict extinction risks in population-based models commonly used in ecology, building on the idea of minimal viable populations \cite{zotero-item-187}. 

Currently, there are still several open questions within viability space decomposition. For starters, it has been argued that given the new organizing structures the theory presents, it should be possible to identify novel bifurcations that track how the map of possible survival outcomes qualitatively shifts as parameters are varied \cite{mcshaffreyDecomposingViabilitySpace2023}. We also must confront how the described manifolds and robustness conditions would behave in systems with chaotic dynamics, which in traditional dynamical systems results in phenomena such as fractal basin boundaries \cite{aguirreFractalStructuresNonlinear2009}. Then there is the issue of scaling. Although we have described viability space in a manner agnostic to the dimensionality of the model system, vast degrees of freedom introduce new technical challenges. As in standard dynamical systems theory, numerically approximating the global manifolds that organize state space is an open research frontier and nontrivial problem \cite{krauskopfSURVEYMETHODSCOMPUTING2005}. More unique to our case, Eqn. 25 establishes that the number of nodes in the agent-graph, and the corresponding number of domains of interaction, scales exponentially with the number of agents. Nonetheless, there are still ways to utilize the theory, even when complete global decompositions become intractable. Just as the concept of a basin boundary is useful even in high dimensional dynamical systems, knowing about these organizing manifolds may permit us to ask new conceptual questions about our models, and engage in more principled explorations of their behavior. Beyond this, our calculations can still tell us about the existence of mortality, ordering, and collapse manifolds and the fate outcomes they separate, even if we do not have a complete picture of their global geometry. There also may be ways to mitigate the exponential scaling of the agent-graph. To start, we may be able to design algorithms that strategically and sequentially look at configurations of agents based on sets we have already found to be stable or vulnerable. Beyond this though, it should be possible to take advantage of the fact that in a lot of agent-based models, the subpopulations are homogeneous in their governing equations and only differ in their state. This could allow us to construct agent-graphs that focus on symmetries across individuals as opposed to having to explicitly represent a domain of interaction for every combination of agents.

Perhaps the greatest takeaway of this paper should be the idea that, if we wish to understand the way states lead to particular fates, we must treat the dynamics of our organisms and the constraints on their viability in a unified manner. In particular, merging the geometric picture of phase space with the geometric concept of a viability region ends up resulting in an object much richer than either of these two concepts in isolation: the viability portrait. While this construction is currently unique to autonomous dynamical systems governed by ordinary differential equations, there is no reason why a similar approach could not be taken to related model formalisms. For example, in stochastic dynamical systems, the viability constraints function as an absorbing boundary that diminishes the probability of observing the agent alive, resulting in the whole space being transiently viable. Nonetheless, it is still possible to ask questions about the distribution of expected lifespans for certain initial conditions. Recent preliminary explorations have shown that in the case of low-to-moderate additive noise, mortality manifolds get blurred into fuzzy bands where the norm of the mean first-passage time gradient is especially steep, separating regions where expected lifespan is relatively robust \cite{mcshaffreyShakingViabilitySpace2025}. While additional analyses need to be carried out in systems with more complex types of noise, this suggests that there can be qualitative alignment between deterministic and stochastic systems with viability constraints at times. Similarly, if time is included as an additional geometric dimension, then similar calculations to viability space decomposition can likely be done for nonautonomous dynamical systems, even if the viability boundary itself is treated as time-varying. Ultimately, the issue of viability is essential to the life sciences, and it will be necessary to develop the theoretical machinery to adequately address it across the range of models and methods that we bring to bear. Developing these insights could allow us to develop rigorous principles for the nature of numerous existential transformations, with implications spanning any field that concerns itself with the integrity of agents.

\begin{acknowledgments}
We thank the many people who have engaged with this work over the course of its development including Eden Forbes, Gabriel Severino, Lindsay Stolting, Denizhan Pak, Joshua Cynamon, Thomas Gaul, Eran Agmon, and James Glazier. We also thank the participants of the Workshop on Theoretical and Experimental Approaches to Goal-Directed Behavior held at the Basque Institute for Applied Mathematics for their questions and feedback. C.M. is supported by the National Science Foundation Graduate Research Fellowship Program under Grant No. 2240777. Any opinions, findings, and conclusions or recommendations expressed in this material are those of the author(s) and do not necessarily reflect the views of the National Science Foundation. 
\end{acknowledgments}

\appendix*

\section{Additional explanation of the behaving cell model}

\begin{figure}[t]
\includegraphics[width=\columnwidth]{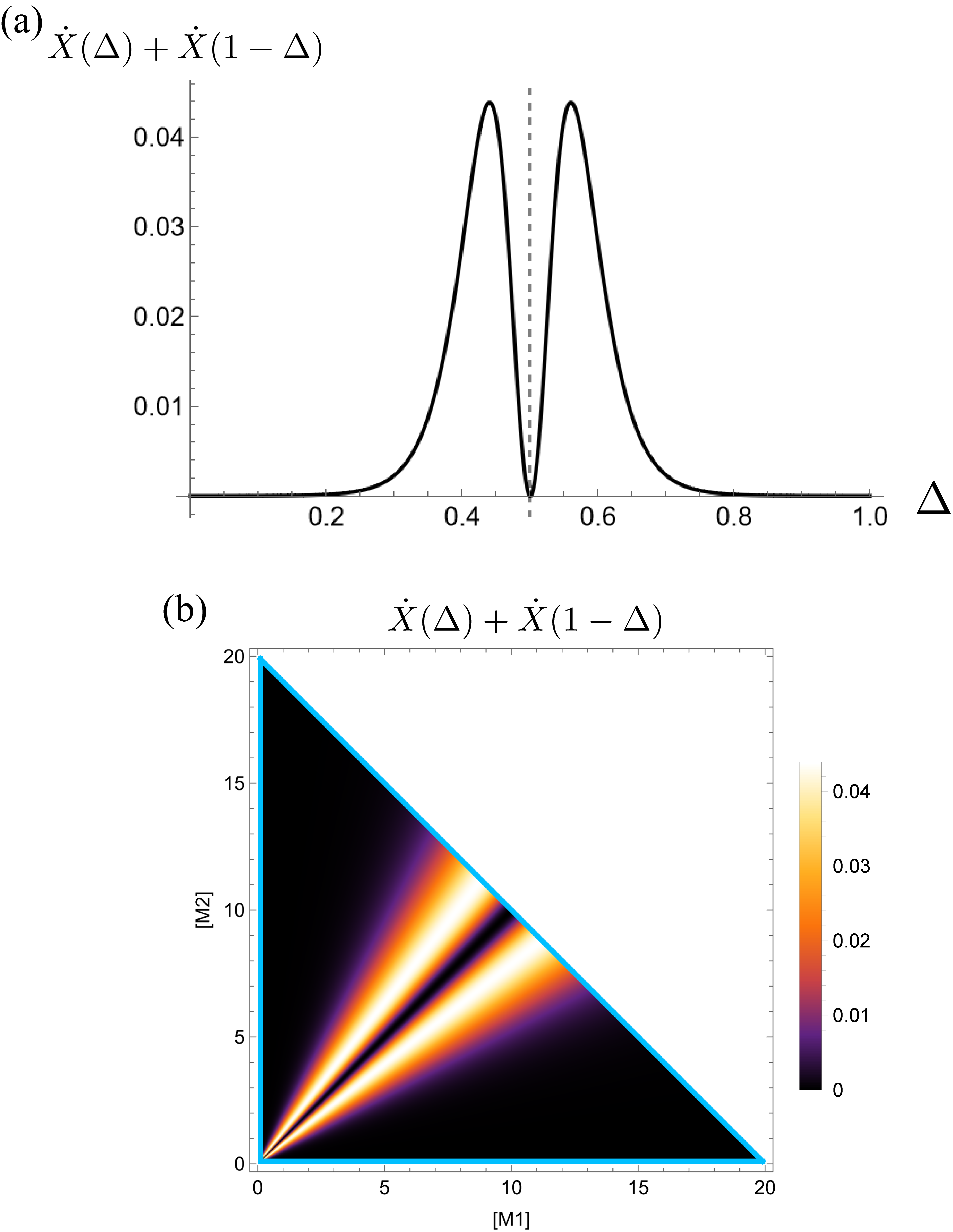}
\caption{\label{fig:epsart} \textbf{(a)} When \(\Delta\) is treated as a variable that can take on values in the range \([0,1]\), the residual calculation from Eqn. A.3 reveals that there is only perfect symmetry when \(\Delta=0.5\), which happens when  \([M_1]\) and \([M_2]\) are exactly equal (gray dashed line). This is why when \(X=0\) the cell does not move and ends up violating both the \(v_\alpha^1\) and \(v_\alpha^2\) constraints simultaneously. Moving away from zero, the residual briefly rises before dipping once again, peaking at approximately \(0.044\). \textbf{(b)} Treating \(\Delta\) as a function of \([M_1]\) and \([M_2]\), we can assess the residual from Eqn. A3 in the physiological ranges permitted by \(\mathcal{V}_\alpha\). This reveals the same peak values as in Fig. 9, but offers more spatial resolution to see how the discrepancy exists in the physiological state space.}
\end{figure}

Back in Section VI B., we mentioned that the behaving cell model is only nearly symmetric in its behavior with respect to the chemical gradients. This is because, while the physiological dimensions and environmental gradients are all in a sense reflections of each other, the behavioral dimension introduces a slight asymmetry due to the basic structure of Hill functions. Consider that if Eqn. 53 had an actual symmetry to it, then swapping the values for \([M_1]\) and \([M_2]\) should always give a velocity that is equal in magnitude but opposite in sign. In our model system, we can essentially break Eqn. 53 into two parts. The first is the comparison between \([M_1]\) and \([M_2]\), \(\Delta\), which exists in a bounded range of \([0,1]\) based on which chemical concentration is dominating:
\begin{equation}
\Delta=\frac{[M_1] - [M_2]}{2([M_1] + [M_2])} + \frac{1}{2}
\end{equation}

\begin{figure*}[t]
\includegraphics[width=\textwidth]{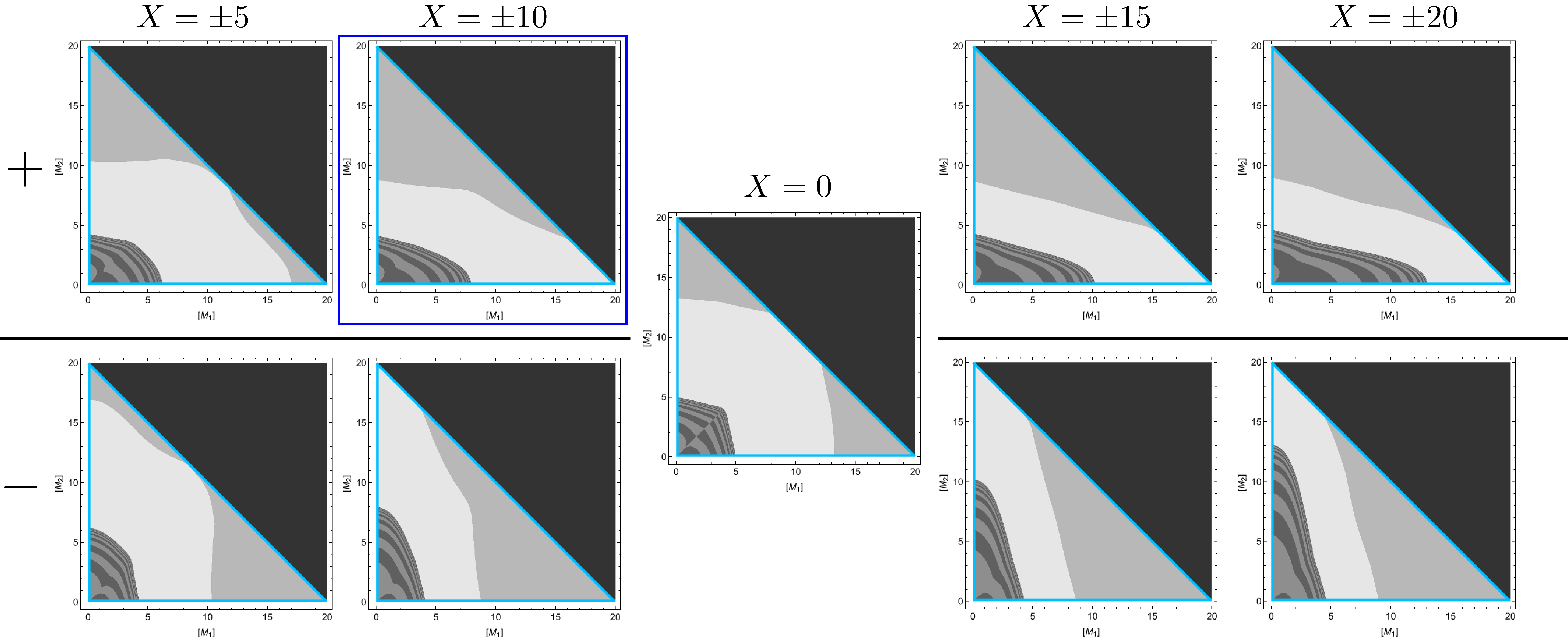}
\caption{\label{fig:epsart} The complete numerical exploration of the behaving cell model, comprising one million simulations per two-dimensional slice.The colors assigned to each fate are the same as in Figs. 5 and 6, and the plane with the blue frame is the same one that was explored in detail in Figs. 6(a), (b), and (c). Despite the slight asymmetry presented in Fig. 9, the positive and negative sides of the environment result in behaviors that are qualitatively identical and quantitatively nearly indistinguishable. The regions carved out by the mortality manifolds \(\mathcal{M}_1\) and \(\mathcal{M}_2\) identified in Figs. 6(g) and (h) and Fig. 7 are not visible in this numerical exploration.}
\end{figure*}

This variable \(\Delta\) is then the input into the shifted inhibitory Hill function that governs the cell's behavior, giving us Eqn. 53:
\begin{equation}
\frac{dX}{dt} = r_\mathrm{spatial} \left( \frac{K_2^{h_2}}{K_2^{h_2} + \Delta^{h_2}} -0.5 \right)
\end{equation}
Now that we can assess the cell's velocity as a function of \(\Delta\), it is possible to see the slight break in symmetry by calculating the residual of the function:
\begin{equation}
\dot{X}(\Delta)+\dot{X}(1-\Delta)
\end{equation}

There are a couple of ways to visualize Eqn. A3. The first is to simply treat \(\Delta\) as a variable that exists in the range \([0,1]\), as shown in Fig. 9(a). This reveals a curve with zero residual when \(\Delta=0.5\), corresponding to \([M_1]\) and \([M_2]\) being equal, followed by two peaks of approximately \(0.044\), and then an approach back to zero residual without ever exactly reaching that value. Another way to visualize the residual is to treat \(\Delta\) as a function of  \([M_1]\) and \([M_2]\) and evaluate across the range of physiological state space within the viability region \(\mathcal{V}_\alpha\), as shown in Fig. 9(b). This reveals the same peak values for the residual as shown in Fig. 9(a), except now we can better see how particular combinations of \([M_1]\) and \([M_2]\) impact the discrepancy as a whole. 

While it is worth acknowledging that the residual is nonzero except where \([M_1]=[M_2]\),  it is also worth noting that the impact of this difference is negligible and has essentially no impact on the qualitative features of the results. For example, Fig. 10 shows the complete numerical exploration of the behaving cell model, with the plane in the dark blue box being the numerical sweep we explored in Figs. 6(a), (b), and (c). Here we can see that as we move away from \(X=0\), the two sides of the environment essentially look like exact reflections of each other, with the \([M_1]\) and \([M_2]\) positions swapping and the \(\mathcal{T}_1\) and \(\mathcal{T}_2\) bands switching the order of their sequence in the lower left of each plane.

Beyond this patterning, the complete numerical sweep of the behaving cells demonstrates several other important features. First off, for any given slice through \(X\), it is still not clear that the layers with the same survival outcome in the bottom left are connected, and that this connection extends across the layers of \(\mathcal{T}_1\) or \(\mathcal{T}_2\) in the other eight slices. Second, the two transiently viable regions carved out by the mortality manifolds \(\mathcal{M}_1\) and \(\mathcal{M}_2\) are nowhere to be seen, emphasizing that even with a high-resolution numerical sweep it is possible to miss fine details in the global structure of fate outcomes. 

If the reverse-time trajectories of the ordering points had been free to continue for infinite time, then the ordering manifold would have resulted in increasingly dense layers without ever terminating, realizing smaller and smaller repetitions of the \(\mathcal{T}_1\)-to-\(\mathcal{T}_2\) patterning in the slices as the trajectories asymptotically approached \(W^s\). The only reason this fractal patterning did not take place is because the reverse-time trajectories eventually terminate at one of the viability boundary segments, representing the earliest viable starting points.

\bibliography{viabdecomp}

\end{document}